\documentclass[runningheads,natbib]{svjour2}
\smartqed  % flush right qed marks, e.g. at end of proof
\usepackage{graphicx}
\def\aj{\rm{AJ}}                   % Astronomical Journal
\def\araa{\rm{ARA\&A}}             % Annual Review of Astron and Astrophys
\def\apj{\rm{ApJ}}                 % Astrophysical Journal
\def\apjl{\rm{ApJ}}                % Astrophysical Journal, Letters
\def\apjs{\rm{ApJS}}               % Astrophysical Journal, Supplement
\def\aap{\rm{A\&A}}                % Astronomy and Astrophysics
\def\mnras{\rm{MNRAS}}             % Monthly Notices of the RAS
\def\prd{\rm{Phys.~Rev.~D}}        % Physical Review D
\def\pasp{\rm{PASP}}               % Publications of the ASP
\def\nat{\rm{Nature}}              % Nature
\def\bain{\rm{Bull.~Astron.~Inst.~Netherlands}} 
\def\physrep{\rm{Phys.~Rep.}}   % Physics Reports
\newcommand{\sm}{\, {\rm M}_{\odot}}

\newcommand{\kms}{\, {\rm km~s}$^{-1}$\,}
\newcommand{\feh}{{\rm [Fe/H]}}
\def\simmore{\mathbin{\lower 3pt\hbox
     {$\rlap{\raise 5pt\hbox{$\char'076$}}\mathchar"7218$}}}   %> or of order
\newcommand{\odeg}{^{\rm o}}
\journalname{Astronomy and Astrophysics Review}
\begin{document}

\title{The stellar halo of the Galaxy%\thanks{Grants or other notes
%about the article that should go on the front page should be
%placed here. General acknowledgments should be placed at the end of the article.}
}

%%%%%%%\subtitle{?\\ If so, write it here}   AH

%\titlerunning{Short form of title}        % if too long for running head

\author{Amina Helmi
}

%\authorrunning{Short form of author list} % if too long for running head

\institute{A. Helmi \at
              Kapteyn Astronomical Institute, \\
	      University of Groningen\\
	      P.O. Box 800, 9700 AV Groningen\\
	      The Netherlands\\
%%%%%              Tel.: +1\\
%%%%%              Fax: +123-45-678910\\
              \email{ahelmi@astro.rug.nl}           %  \\
%             \emph{Present address:} of F. Author  %  if needed
}

\date{Received: date}
% The correct date will be entered by the editor

\maketitle

\begin{abstract}
Stellar halos may hold some of the best preserved fossils of the
formation history of galaxies. They are a natural product of the
merging processes that probably take place during the assembly of a
galaxy, and hence may well be the most ubiquitous component of
galaxies, independently of their Hubble type. This review focuses on
our current understanding of the spatial structure, the kinematics and
chemistry of halo stars in the Milky Way. In recent years, we have
experienced a change in paradigm thanks to the discovery of large
amounts of substructure, especially in the outer halo. I discuss the
implications of the currently available observational constraints and
fold them into several possible formation scenarios. Unraveling the
formation of the Galactic halo will be possible in the near future
through a combination of large wide field photometric and
spectroscopic surveys, and especially in the era of Gaia.
\keywords{Galaxy: halo \and Galaxy: formation \and Galaxy: evolution
\and Galaxy: kinematics and dynamics }
\end{abstract}

\section{Introduction}
\label{intro}

A word of caution is necessary before we start our journey through the
stellar halo of the Galaxy. While working on this article, I could not
help but wonder whether this was a good time to write a review on the
stellar halo \citep[see, for example, previous excellent reviews
by][]{gwk,majewski-rev,bhf,freeman}.  The advances in the recent past
have been enormous thanks to many (ongoing) large wide field
photometric and spectroscopic surveys.  In the past two years, almost
every month a new satellite galaxy or a new stream was
discovered. This in itself may be a justification for writing a
review, but it also may lead to an article that is rapidly out of
date.

On the other hand, this of course, is only a sign of the field's
health and its great promise for the young generations of scientists
hoping to unravel the formation of the Galaxy.  The reader is hereby
warned that our understanding of the Galactic halo is still evolving
significantly. In an attempt to avoid the (almost unavoidable) risk of
becoming quickly outdated, the author has decided to review those
properties of the stellar halo that (hopefully) provide the most
direct clues to its evolutionary path.

The stellar halo is arguably the component that contains the most
useful information about the evolutionary history of the Galaxy. This
is because the most metal-poor stars in the Galaxy and possibly some
of the oldest ones are found here. It therefore provides us with a
picture of the Milky Way in its early stages of evolution. Low-mass
stars live for much longer than the present age of the Universe, and
so retain in their atmospheres a record of the chemical elements of
the environment in which they were born. The very metal-poor halo
stars are thus fossils, whose chemical abundance and motions contain
information of their sites of origin. As \citet{els} nicely put it:
``{\sl ... a study of these subsystems allows us partially to
reconstruct the Galactic past because the time required for stars in
the Galactic system to exchange their energies and momenta is very
long compared with the age of the Galaxy. Hence knowledge of the
present energy and momenta of individual objects tells us something of
the initial dynamic conditions under which they were formed}''.

Despite its crucial importance, our knowledge of the stellar halo is
fragmentary. This is at least partly driven by the paucity of
halo stars (near the Sun roughly only one star in a thousand belongs
to the halo, as discussed in Section~\ref{sec:norm}). This implies
that generally a set of suitable (and preferably unbiased) selection
criteria are used to find these precious fossils, as we shall see
later in this review.

The study of the formation of the Galaxy is particularly relevant at
this point in time. There is a cosmological framework in place, the
$\Lambda$ cold dark matter (CDM) paradigm, which broadly describes
from first principles how structures in the Universe have evolved
\citep[e.g.][and references therein]{springel}. In this model,
galaxies form hierarchically, through the amalgamation of smaller
proto-systems. Despite the many successes of this theory, particularly
on large scales, we do not yet understand in detail how galaxies
form. Many of the shortcomings of this model have become evident
through comparisons to the properties of our Galaxy and its nearest
neighbors \citep[e.g. the ``missing satellites''
problem,][]{klypin99,moore99}. Therefore, it becomes clear that studies
of the Galaxy and its components are crucial for a proper and complete
understanding of galaxy evolution in the Universe.

This review describes the various characteristics of the stellar halo
in the hope that these will aid us in our understanding of the
formation of the Galaxy. We start by discussing the spatial structure
and global kinematics in Section~\ref{sec:struct}. This approach is rather
classical, since we shall discuss separately the presence of substructure in
Section~\ref{sec:subst}. In principle, substructure can strongly affect
the determination of the density profile or of the shape of the stellar
halo, e.g. by introducing overdensities which may not always be easy
to identify as such \citep[e.g.][]{newberg-yanny,xu07,Bell}. Given our
current understanding of the Galactic halo, such substructures may
well dominate at large radii, but at small distances (i.e. the stellar
halo near and interior to the Solar circle), the mixing timescales are
sufficiently short that even if all of the stellar halo had been
formed through a sequence of dissipationless mergers it would be
completely smooth spatially \citep[see][]{hw99}.

In Section \ref{sec:chemistry} we describe our current knowledge of
the metallicity distribution of halo stars, and particularly focus on
what it reveals about star formation in the early Universe. We also
discuss what chemical abundance patterns tell us about the process of
formation of the Galaxy. Not only halo field stars are interesting,
but also globular clusters and satellite galaxies give us insight into
the characteristics and evolution of the Milky Way halo. These are
discussed in Section \ref{sec:satellites}.

In Section \ref{sec:formation} we discuss formation scenarios for the
stellar halo taking into account most (if not all) observational
constraints presented in earlier sections.  Recent technological
advances have allowed us to begin to probe halos in other
galaxies. Their characteristics and a comparison to our own stellar
halo are briefly mentioned in Section~\ref{sec:ext-halos}.  Finally, we
provide a brief summary and outlook in Section \ref{sec:concl}.

\section{Spatial structure and global kinematics}
\label{sec:struct}

\subsection{Star counts: density profile and shape}
\label{sec:star-counts}

Star counts have been used in astronomy for over a century to derive
the structure of our Galaxy \citep[and even of the Universe,
e.g.][]{kapteynvanRijn,shapley}. They have led to the conclusion
that our Galaxy has several physically distinct components, namely the
thin and thick disk, the stellar halo and the bulge/bar
\citep[e.g.][]{bahcall,gilmore-reid}.

The structure of the stellar halo is intimately linked to how the
Galaxy formed \citep{freeman}. Different scenarios predict different
shapes and correlations with properties such as age, metallicity,
etc., as we shall discuss later. In the simplest case, the structure
of the stellar halo could also give us insight into the structure of
the dark matter halo. For example, if a significant fraction of the
dark matter had been baryonic \citep[composed by MACHOs,][]{pac}, then
the stellar and dark halos would presumably be
indistinguishable. However, this scenario appears unlikely, as the
microlensing surveys are unable to assign more than 8\% of the matter
to compact dark objects \citep[][although see \cite{machos} for a
different while earlier result]{eros}.

In modern cosmological models (namely, cold dark matter dominated), it
is now clear that the structure of the stellar and dark halo do not
necessarily follow one another. The dark halo is much more extended,
and probably follows an NFW profile \citep{nfw}. Its half-mass radius
lies around 150 kpc \citep{klypin,gb05,gb05err}, and therefore its
structure is probably related to the present-day nearby large scale
tidal field \citep{nas}. On the other hand, the stellar halo is highly
concentrated (e.g. the effective radius as traced by the halo globular
clusters is well within the Solar circle, \citet{fw}), a property that
likely reflects a very early formation epoch\footnote{This refers to
the inner stellar halo; as we shall show in Section~\ref{sec:subst} it
is clear that the outer stellar halo is still forming.}
\citep{moore2001,hws}.

In the early years, star counts were based on photographic images, and
were usually restricted to just a few fields. This was also the basis
for the pioneering \citet{bahcall} model of the Galaxy. These authors
took the approach that our Galaxy is ``just'' like any other galaxy,
and so it should be possible to reproduce the star counts using models
that fit the light distribution of external galaxies. For example,
they proposed the halo followed a ``de Vaucouleurs'' spherical
profile, which is characteristic for elliptical galaxies (although it
is now known that not all elliptical galaxies are well-fit by this
function; furthermore \cite{morrison93} showed that a power-law $\rho
\propto r^n$ with $n \sim -3.5$ as proposed by \cite{harris} on the
basis of halo globular clusters, provides a better fit for the stellar
halo than an $R^{1/4}$ law).

Already in early studies \citep{kinman66} there was evidence that the
stellar halo's shape varied with distance from the Galactic
center. These studies favored a halo that was rounder in the outskirts
(minor-to-major axis ratio $q\sim1$) and more flattened in the inner
parts \citep[$q\sim0.5$;][]{Schmidt,hartwick87,preston91}. Typically
RR~Lyrae and blue horizontal branch stars (BHB) were used as
tracers. Even today these stars are commonly used to isolate samples
of true halo stars. Such stars are only found in old populations, and
they are relatively bright so they can be observed with modest
telescopes at distances as large as 100 kpc from the Galactic center
\citep[e.g.][]{clewley2}.

The density profile of the stellar halo is often parametrized in a
principal axis cartesian coordinate system as
\begin{equation}
\rho(x,y,z) = \rho_0 \frac{\bigg(x^2 + \displaystyle \frac{y^2}{p^2} +
\displaystyle \frac{z^2}{q^2} + a^2\bigg)^{n/2}}{r_0^n}
\label{eq:rho}
\end{equation}
where $n$ is the power-law exponent (and because of the finite extent
of the Galaxy, it is negative), and $q$ and $p$ are the minor- and
intermediate-to-major axis ratios (in the axisymmetric case, $p=1$).
The scale radius $a$ is often neglected because of the rather
concentrated nature of the stellar halo \citep[although see for
example][]{phleps,bica}. Finally, $\rho_0$ is the stellar halo density
at a given radius $r_0$, which is generally taken to be the solar
radius for obvious reasons (and referred to as the local
normalization). As we shall see below, this parametrization is likely
to be an oversimplification because the shape as well as the exponent
seem to vary with distance from the Galactic center.

\subsubsection{Modern pencil beam surveys}

\begin{figure}
\includegraphics[width=0.3\textwidth,angle=270]{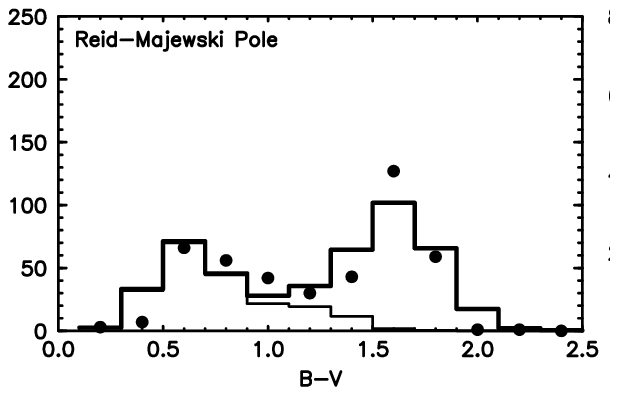}
\includegraphics[width=0.3\textwidth,angle=270]{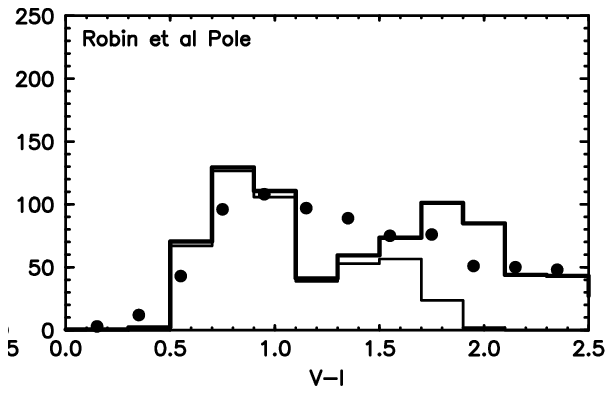}
\begin{center}
\vspace*{0.3cm}
\includegraphics[width=0.5\textwidth,clip]{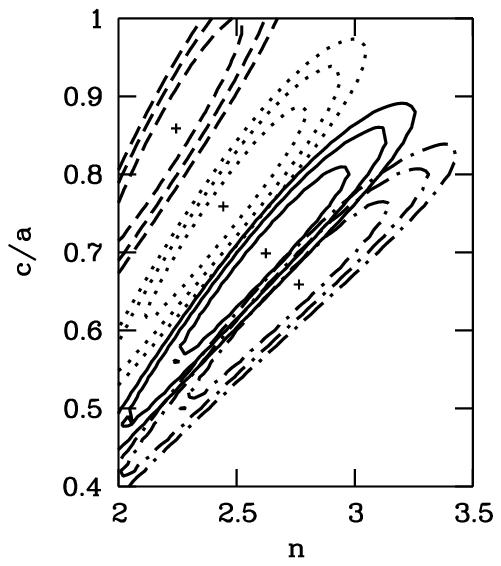}
\end{center}
\caption{{\it Top panels}: Color distributions of two fields used by
\citet{robin00} to constrain the shape of the stellar halo. The dots
correspond to the true number of observed stars in each color bin,
while the heavy solid lines are the number of stars predicted by the
model assuming a power law index of $-2.44$, a flattening of 0.76 and
a local density of 0.75$\times \rho_0$ for the stellar halo. Thin
lines show the contribution of the halo alone. The red peak
corresponds to nearby M dwarfs, while the blue peak indicates the
location of the halo and thick disk turn-off points. {\it Bottom
panel}: Iso-likelihood contours at 1, 2 and 3 sigmas in the plane
(power law index, flattening) of the spheroid population, coded by the
local density, going from 1 (solid), 0.75 (dotted); 0.50 (dashed) and
1.25 (dash-dotted)$\times\rho_0$. This diagram evidences the strong
covariance between the shape and power-law index of the stellar
halo. [From \cite{robin00}. Courtesy of Annie Robin. Reproduced with
permission of A\&A.]}

\label{fig:halo-deg}
\end{figure}

\citet{robin00} have used a large combined set of deep star counts at
high and intermediate galactic latitudes, and found a power-law index
$n=-2.44$ and flattening $q=0.76$.  \citet{siegel} observed seven
Kapteyn selected areas (14.9~deg$^2$) (hence ensuring uniform
photometry) and preferred $n = -2.75$ and $q=0.6$. There are many
delicate points in these analyses. For example, the counts also
include stars from the thin and thick disks, which implies that the
fits should also allow for the (self-consistent) modeling of these
components (see top panels in Figure \ref{fig:halo-deg}). Furthermore
there is a degeneracy in the modeling of the stellar halo in the sense
that a less flattened halo requires a steeper density profile (bottom
panel of Figure \ref{fig:halo-deg}). However, it is worth bearing in
mind that in both studies the fits obtained are not perfect, and
systematic differences between models and data are apparent,
especially at the faint end (see also Figure 6 in \cite{robin07}). In
particular, there are more stars towards the anticenter than
predicted, and there are less stars observed towards the center. This
is similar to what \citet{spagh-1} find using turn-off halo stars
(i.e. those with $(B-V)\sim0.38$), who prefer $n=-3$ and $q =
0.6$. These authors note that a two-component halo is unlikely to
solve these problems, because it is incapable of enhancing the number
counts towards the anticenter (a rounder halo would in fact tend to
lower the counts). Some proposed solutions are that the thick disk
scale height varies with radius, or that the outer halo is
lumpy. Either may well be possible. For example, fields at Galactic
latitudes $|b| < 25\deg$ are now known to contain significant
substructure towards the anticenter
\citep{newberg,ibata,grillmair}. More recently CADIS (Calar Alto Deep
Imaging Survey; \cite{phleps}) mapped 1/30 deg$^2$ in the magnitude
range $15.5 \le R \le 23$, and found $n=-2.5$ if $q = 0.6$.

\subsubsection{Panoramic surveys}

The advent of large galaxy surveys has been extraordinarily beneficial
for Galactic astronomy. Although initially conceived to be a by
product, multi-color deep wide field surveys have yielded completely
new views of the Galactic halo. Some of the most spectacular results
come from the Sloan Digital Sky Survey (SDSS\footnote{\sf
http://www.sdss.org}). The first studies on the Early Data Release
(EDR; 100 deg$^2$ down to $r \sim 20$) were carried out by
\citet{ivezic} using RR~Lyrae; \citet{yanny00} using A-stars (BHBs and
blue stragglers) and \citet{chen} with stars of halo turn-off
color. In agreement with previous work using the same technique (see
previous section), \citet{chen} measured $n = -2.3/2.5$ and $q =
0.6/0.5$ depending on the local normalization of the halo. However,
\citet{yanny00} estimated $n = -3.2 \pm 0.3$ and $q = 0.65$ (after
masking out the overdensities due to the Sagittarius streams), in
accordance with \citet{ivezic} who found $n\sim -3$. Note that the
latter two works employ tracers that probe considerably larger
distances \citep[particularly in comparison to the halo turn-off
stars, which are intrinsically much fainter: $M_{V_{\rm TO}} \sim 4$ versus
$M_{\rm V_{HB}} \sim 0.6$;][]{binney-merr}.

Using RR~Lyrae stars, the QUEST survey \citep{vivas} find $n=-3.1 \pm
0.1$ (and that steeper profiles produce worse fits), and $q$ varying
with radius as proposed by \citet{preston91}.

The SDSS has now imaged 8000 deg$^2$ of the sky yielding a truly
panoramic view of the Galaxy \citep{dr5}. However, instead of
simplifying the picture and providing definitive answers, it has shown
that the halo is far more complex than envisioned some ten years
ago. For example, \citet{newberg-yanny} find that there are
significant asymmetries with respect to the line connecting the
galactic center to the anticenter in the number counts of halo turnoff
stars. They postulate this could be due to a triaxial halo.  In
concert with this, \citet{xu} find an excess of stars towards
$[l,b]=(270\odeg,60\odeg)$ compared to $[l,b]=(90\odeg,60\odeg)$.

\begin{figure}
\includegraphics[width=0.5\textwidth]{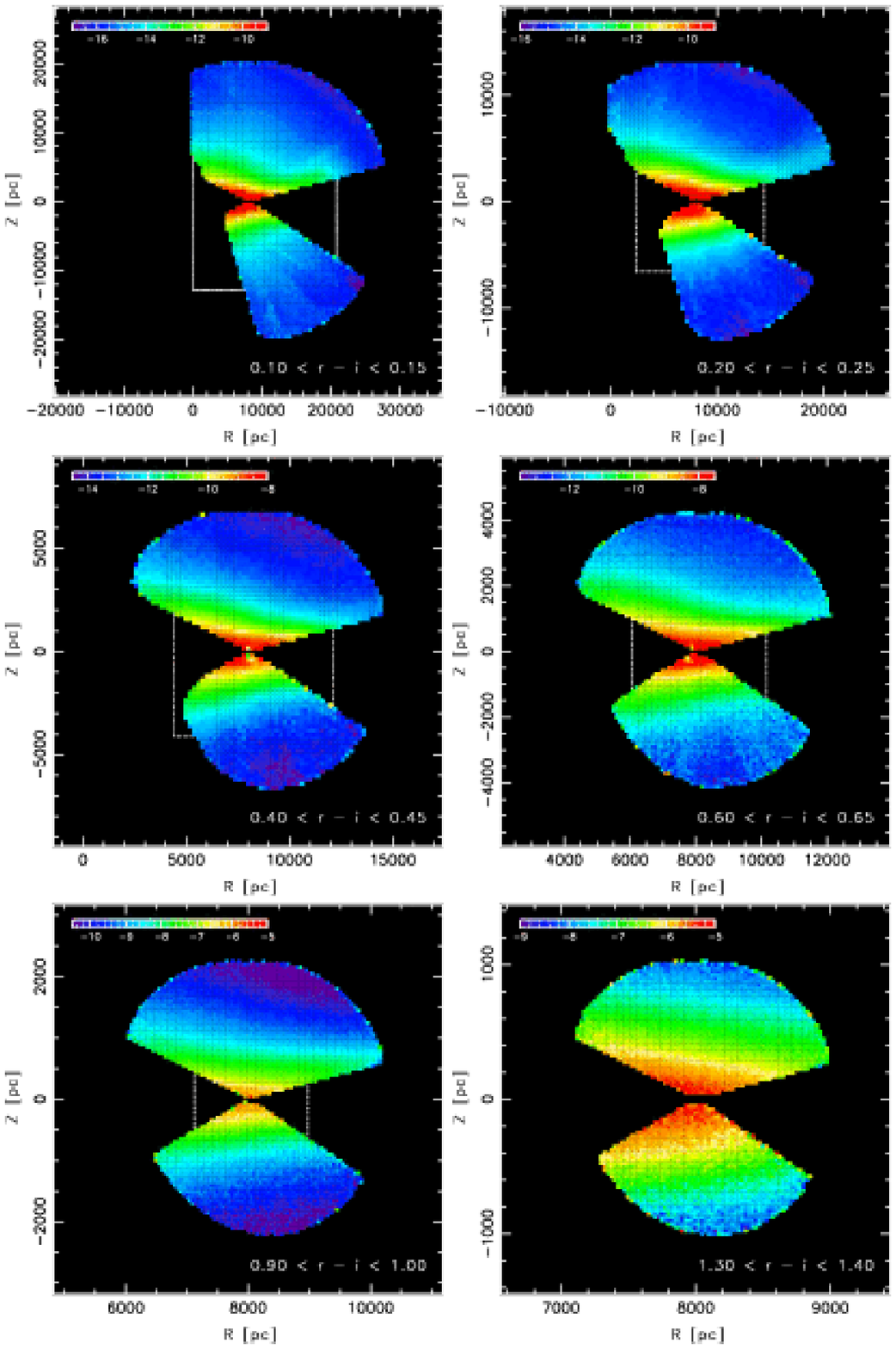}
\includegraphics[width=0.5\textwidth]{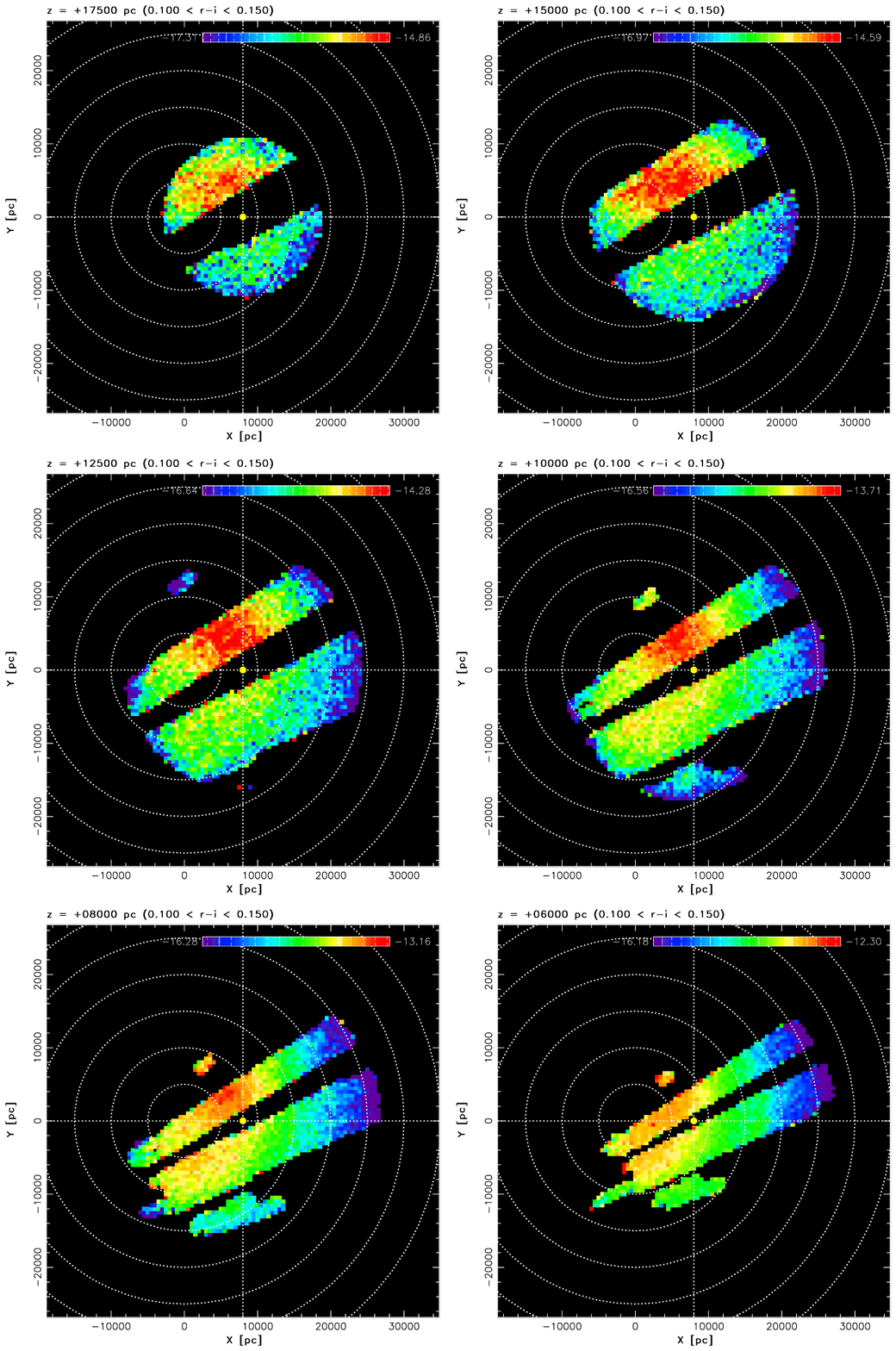}
\caption{{\it Left panels}: Stellar number density as a function of
$(R,z)$ for different $r-i $ bins using SDSS photometric parallaxes
(the distance scale greatly varies from panel to panel). Each white
dotted rectangle denotes the bounding box of the region containing the
data on the following panel. Note the large overdensities particularly
in the top panels. {\it Right panels}: The stellar number density for
the same color bin as in the top left panel ($0.10 < r-i < 0.15$,
hence preferentially selecting stars near the halo turn-off point),
shown here for cross-sections parallel to the Galactic plane, as a
function of the distance from the plane (from 16 kpc at the top left
to 6 kpc in the bottom right). The origin marks the Sun's position,
while the circles aid in visualizing departures from axial symmetry of
the Galaxy. Note the strong asymmetry with respect to the $y=0$ line,
which is due to the Virgo overdensity. [From \cite{juric}. Courtesy of
Mario Juric. Reproduced with permission of the AAS.]}
\label{fig:juric}
\end{figure}

\citet{juric} have derived photometric parallaxes\footnote{assuming a
``mean'' photometric parallax relation, as function of color index,
for all metallicities.} for 48 million stars in the SDSS catalog and
attempted full modeling of the various Galactic components. These
authors find that it is possible to measure the properties of the thin
and thick disk relatively robustly (provided large portions of the
celestial sphere have been observed). However they acknowledge that
determining the properties of the halo is more difficult, partly
because of the presence of significant clumps and overdensities, as
shown in Figure \ref{fig:juric}. The halo power law index $n$ is
poorly constrained by the data, but an oblate halo with $q = 0.5$
seems to be favored, although no good fits are obtained at the faint
end.  Amongst various structures, \citet{juric} have detected a huge
overdensity towards the constellation of Virgo at $l=270\odeg$ and
$b=60\odeg$ (the same region that cannot be reproduced by smooth
axisymmetric models).

The SDSS is focused on the northern galactic cap. This implies that
the nature of the overdensities (global versus local) may be difficult
to establish without good coverage of the southern hemisphere. In
particular, if the halo were triaxial, one would still expect (in the
most reasonable case) symmetry with respect to the $z=0$
plane. \citet{xu07} carried out this exercise and compared star counts
towards the North Galactic Pole (NGP) from SDSS and towards the South
Galactic Pole (SGP) using SuperCOSMOS\footnote{\sf
http://www-wfau.roe.ac.uk/sss/} photographic data \citep{hambly}. They
found that there is an asymmetry between the north and southern cap of
approximately 17\% in the $B$-band. Therefore, they concluded that the
Virgo overdensity is likely a foreign component of the stellar halo
(see also Section~\ref{sec:subst}). Their modeling, after subtracting
the overdensities, favors an axisymmetric halo with parameters
$n=-2.8$ and $q = 0.7$.

This last example shows particularly well the need for large sky
coverage and deep photometry to establish the structure of the stellar
halo of the Galaxy. This is the only way to break the known
degeneracies affecting the modeling of number counts from restricted
pencil beam surveys. Nevertheless, given the problems in reproducing
star counts at the faint end and the discovery of large overdensities
on the sky, it is essential that the presence of substructure is taken
into account in the models.

At this point in time, it transpires that the inner halo may well be
axisymmetric and flattened ($q \sim 0.6$), while the outer halo
(traced by e.g. RR~Lyrae stars and very deep counts as in
\cite{robin07}) may be more spherical (and steeper). It would be
interesting to explore whether this is due to different formation paths
for the inner and outer halo (e.g. dissipative versus non-dissipative
formation, see Section \ref{sec:formation-global}).  The gravitational
pull of the stellar disk (and the adiabatic contraction of the halo)
may also be (partly) responsible for this dichotomy, see
e.g. \citet{binney-may,chiba-beers01}. In principle, it is also possible
that the dual shape of the stellar halo reflects a change in the
number of progenitors and their orbital characteristics with distance
from the Galactic center.

\subsection{Normalization}
\label{sec:norm}

The local normalization (density of the stellar halo at the position
of the Sun) has been determined both locally as well as from deep
pencil beam surveys. \cite{morrison93} isolated local halo giants from
an objective-prism survey of metal weak stars, and derived a halo
number density expressed in terms of halo-to-disk ratio of 1:1200 for
K giants ($M_V \le 0.5$), and 1:850 for dwarf stars with magnitudes $5
\le M_V \le 8$. The latter is comparable to the estimate obtained by
\citet{fuchs-jahreiss} using nearby M dwarfs (within 25 pc), who
measured $\rho_0 = 1.5 \times 10^{-4}\sm/$pc$^3$ \citep[see
also][]{digby}. On the other hand, \citet{flynn} used very deep HST
counts of M dwarfs and obtained a factor four smaller number of red
dwarfs, in agreement with \citet{gould98} who find $\rho_0 \sim 6.4
\times 10^{-5} \sm/$pc$^3$ using subdwarfs in 53 fields observed with
the Wide Field Planetary Camera on the Hubble Space Telescope
(HST). Attempts to measure the normalization using BHB stars are more
tricky because of the variation of the properties of the horizontal
branch with age and metallicity. This implies that the transformation
from number of BHBs to $M_\odot$ is non-trivial. Note that the
measurement of $\rho_0$ relies on an accurate determination of the
luminosity function (and hence on the slope of the initial mass
function), but possibly the largest source of discrepancy is the
varying shape of the halo with radius.  For example, the deep counts
using HST explore larger distances, and hence presumably also the
rounder part of the stellar halo, and therefore necessarily favor a
rather low local normalization \citep{fuchs-jahreiss}.

\subsection{Kinematics}
\label{sec:kinematics}

Obtaining space velocities for halo stars is no doubt more laborious
than mapping their sky distribution. Although radial velocities are
relatively straightforward to obtain for a small number of objects
even for rather faint stars (e.g.  $V \sim 19$ take less than one
hour-long integrations with 4m-8m class telescopes), measuring their
tangential velocities is only possible if both the proper motion and
distance are known. While estimates of the latter may be obtained
through photometric or spectroscopic parallaxes, proper motions from
the ground require a large time baseline (of the order of 100 years)
to produce milliarcsecond accuracy \citep[see, e.g.][and references
therein]{beers-cat}. Therefore, and for many other reasons
\citep{lindegren}, astrometric space missions with wide field angle
capabilities are an absolute necessity.

Presently the catalogs of halo stars with relatively accurate 3D
kinematics are somewhat small, and typically sample a few kpc around
the Sun. They have been constructed using proper motions mainly from
Hipparcos, and radial velocities from other surveys. Velocity errors
are in the range 20 -- 30\kms. Currently the largest such catalog
available is that by \cite{beers-cat}, with approximately one thousand
nearby stars with $\feh < -0.6$~dex.

\subsubsection{Local halo}

Local samples of halo stars show a small amount of prograde rotation
$V_\phi \sim$ 30--50\kms
\citep{carney96,layden98,chiba-beers,morrison07}. The velocity
ellipsoid is roughly aligned with a cylindrical coordinate system,
having dispersions of $(\sigma_R,\sigma_\phi,\sigma_z)=(141 \pm 11,
106 \pm 9, 94 \pm 8)$\kms \citep{chiba-beers}. The stellar halo is
therefore supported by random motions, and its flattened shape is
consistent with the degree of velocity anisotropy observed \citep[see
Section~\ref{sec:shape-ellips}, also][]{sl-c}.

The velocity distribution of halo stars near the Sun appears to be
fairly smooth, with only a small degree of substructure (see
Section~\ref{sec:subst} for more details). There is no strong correlation
of the rotational velocity $V_\phi$ with metallicity. However, at
large distances above the Galactic plane ($\sim 3 - 5$ kpc),
the halo becomes slightly retrograde (\citet{majewski96,carney96}, see
also \cite{kinman07}), and its mean metallicity is lower
\citep{carollo,morrison07}.

\subsubsection{Global properties}

\begin{figure}
\includegraphics[width=0.5\textwidth]{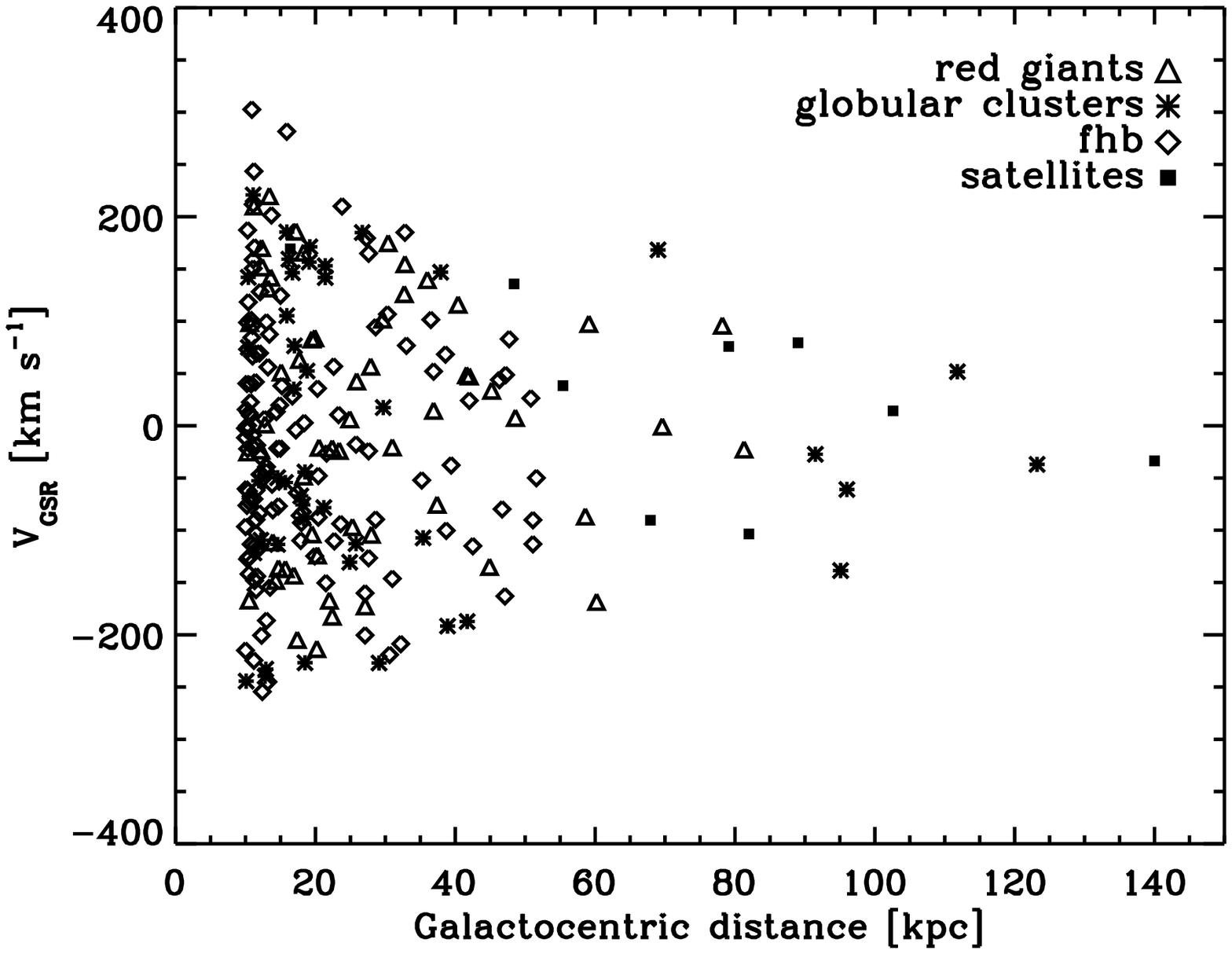}
\includegraphics[width=0.5\textwidth]{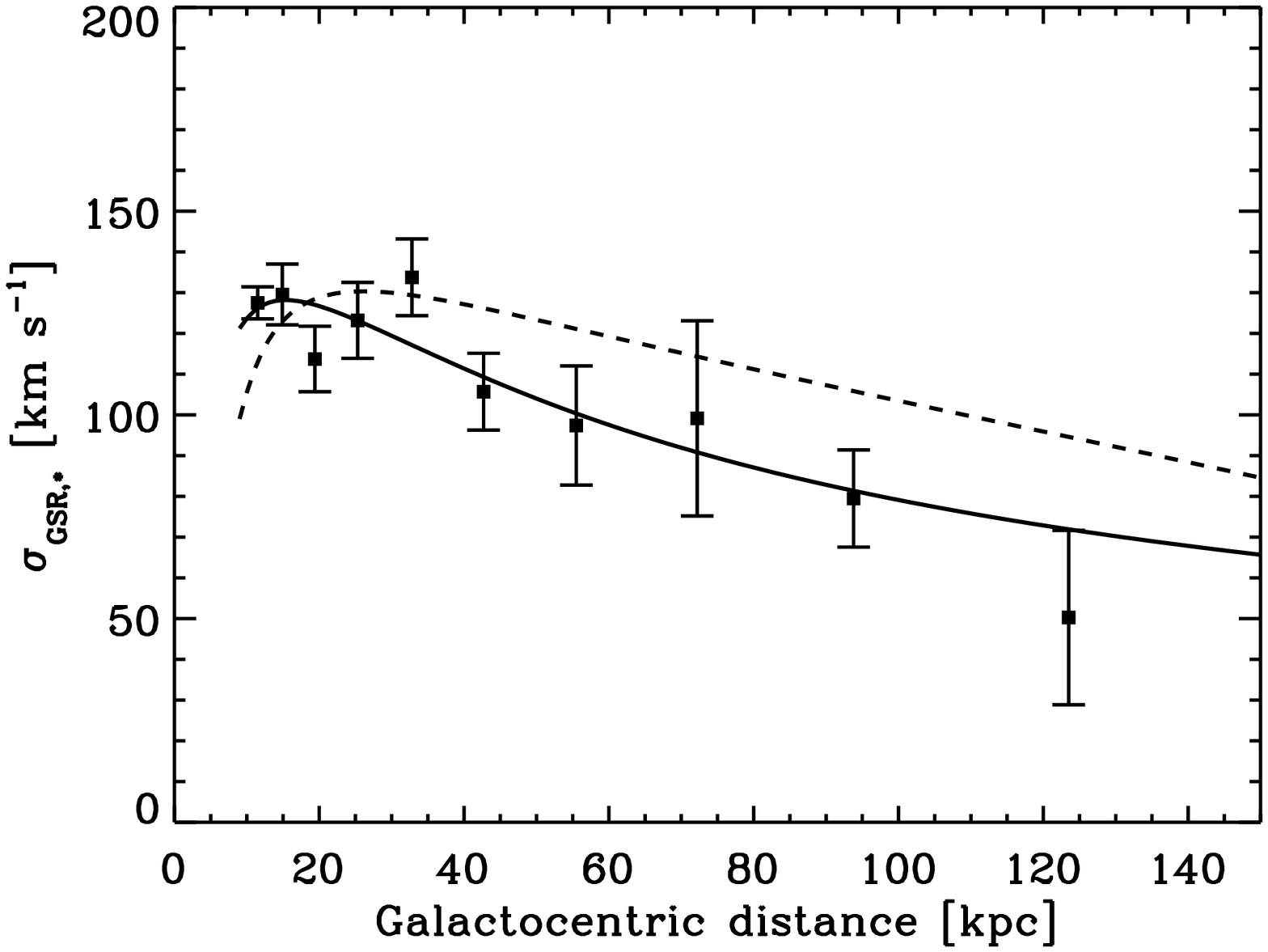}
\caption{Radial velocities in the Galactic Standard of Rest $V_{\rm
GSR}$ as function of distance from the Galactic center ({\it left});
and Galactocentric radial velocity dispersion of the Milky Way stellar
halo ({\it right}). The curves show the best-fit models obtained
assuming a truncated flat dark matter halo of mass $5^{+2.5}_{-1.7}
\times 10^{11} \sm$ (solid curve) and an NFW profile of mass
$9.4^{+1.4}_{-0.9}\times 10^{11} \sm$ and c= 18 (dotted curve), with
constant anisotropy. The strong decline suggests that for an NFW
profile to provide a better fit either the anisotropy must change with
radius, or the stellar halo must have an edge. [From
\cite{gb05,gb05err}. Courtesy of Giuseppina Battaglia. Reproduced with
permission of Wiley-Blackwell Publishing.]}
\label{fig:spag-giusy}
\end{figure}

Figure \ref{fig:spag-giusy} shows the line of sight velocities of a
sample of 240 objects in the Galactic halo \citep{gb05,gb05err} as
function of distance from the Galactic center. This sample includes K
giant stars from the Spaghetti survey, field horizontal branch stars,
globular clusters and dwarf galaxies. On the right-hand side panel,
the radial velocity dispersion is plotted. This quantity shows an
almost constant value of 120\kms out to 30 kpc and then continuously
declines down to 50\kms at about 120 kpc (see \cite{sl-sigma} for an
earlier similar conclusion from a smaller and less distant
sample). This fall-off puts important constraints on the density
profile and total mass of the dark matter halo of the Milky Way. In
particular, under the assumption of a power-law density for the
stellar halo and a constant velocity anisotropy, an isothermal profile
can be ruled out, while both a dark halo following a truncated flat
model \citep{we99} of mass $5^{+2.5}_{-1.7} \times 10^{11} \sm$ and an
NFW profile of mass $9.4^{+1.4}_{-0.9}\times 10^{11} \sm$ and
concentration $c= 18$ are consistent with the data \citep[see
also][]{moore06,abadi06}.

The stellar halo is not completely smooth kinematically (see Section
\ref{sec:subst}). Evidence of halo moving groups was presented by
\cite{rf85} towards the South Galactic Pole using radial velocities
for K giants, by \cite{majewski96} towards the North Galactic Pole
based on proper motions, and more recently by e.g. \cite{clewley},
\cite{sirko} and \cite{vivas} using horizontal branch stars' radial
velocities. We discuss these substructures in more detail in
Section~\ref{sec:subst}.

\subsubsection{A small digression on the relation between the shape of 
the stellar halo and its velocity ellipsoid in the Solar neighborhood}
\label{sec:shape-ellips}

As described above, the stellar halo does not show significant
rotation.  This implies that its flattened shape has to be attributed
to anisotropy in the velocity ellipsoid.\footnote{A similar argument
is used to explain the shape of elliptical galaxies, see \citet{bt}.}

We will now derive the expected relation between the flattening of the
stellar halo and the degree of anisotropy using the virial theorem.
In steady state, the virial theorem in tensor form may be expressed as
\begin{equation}
2 T_{jk} + \Pi_{jk} + W_{jk} = 0
\end{equation}
where 
\begin{eqnarray}
T_{jk} &=& \frac{1}{2} \int \,d^3x\, \rho \,\bar{v}_j\bar{v}_k
\nonumber \\ \Pi_{jk} &=& \int \,d^3x\, \rho \,\sigma^2_{jk} \nonumber
\\ W_{jk} &=& \int \,d^3x\, \rho \, x_k \frac{\partial \Phi}{\partial
x_j} \nonumber
\end{eqnarray}
with $\rho({\bf x})$ the density profile of the system, $\bar{v}_i$
the mean velocity along the $i$-axis, $\sigma^2_{jk}$ its velocity
tensor, and $\Phi({\bf x})$ the gravitational potential of the
dominant mass component. Note that in the case of the stellar halo,
the density and the gravitational potential are not related through
Poisson's equation; the stellar halo is a tracer population embedded
in an underlying potential whose main contributor is the dark matter
halo.

If we assume that the dark halo is spherically symmetric and that the
stellar halo is axisymmetric, then $W_{xx} = W_{yy}$ and $W_{jk} = 0$
for $j \ne k$. Similar arguments can be used to show that all
equations are automatically satisfied for $j\ne k$, while the only two
independent equations that remain are
\begin{eqnarray}
2 T_{xx} + \Pi_{xx} + W_{xx} = 0 \nonumber \\
2 T_{zz} + \Pi_{zz} + W_{zz} = 0.\! \nonumber
\end{eqnarray}

Because the streaming (mean) motion of the stellar halo is negligible
compared to the random motions, we may set $T_{xx} = T_{zz} = 0$. In
this case, 
\begin{equation}
\frac{\Pi_{xx}}{\Pi_{zz}} = \frac{W_{xx}}{W_{zz}}
\label{eq:ratios}
\end{equation}
which states that shape of the mass-weighted velocity dispersion
tensor depends on the shape of the (moments of the) mass-weighted
gravitational potential.  Equation (\ref{eq:ratios}) is equivalent to
\begin{equation}
\frac{\int d^3x\, \rho(R,z) \,\sigma^2_{x}}{\int d^3x\, \rho(R,z)
 \,\sigma^2_{z}} = \frac{ \int \,d^3x\, \rho(R,z) \, \displaystyle
 \frac{x^2}{r} \displaystyle \frac{d \Phi}{d\,r}}{\int \,d^3x\, \rho(R,z) \,
 \displaystyle \frac{z^2}{r} \displaystyle \frac{d \Phi}{d\,r}}.
\label{eq:final_form}
\end{equation}
Note that $\sigma_x \ne \sigma_z$ if the density distribution of the
tracer is axisymmetric. For simplicity we will assume
\begin{enumerate}
\item The shape of the velocity ellipsoid does not depend on location
\begin{equation}
\sigma^2_x/\sigma^2_z = {\rm constant}.
\end{equation}

\item A logarithmic potential for the dark halo
\begin{equation}
\Phi(r) = v_c^2 \ln (r^2 + d^2).
\end{equation}

\item The density profile of the stellar halo follows a power-law, is
axisymmetric and has flattening $q$. Using Eq.~\ref{eq:rho} we obtain
\begin{equation}
\rho(R,z) = \rho_0 (m/m_0)^{n}, \qquad m^2 = R^2 + z^2/q^2.
\end{equation}
\end{enumerate}

Under these conditions, the left-handside of Eq.~(\ref{eq:final_form})
reduces to a constant. The goal now is to evaluate the right-handside
expression and compare it to the numerical value derived
observationally for this constant. 

Let us consider the case in which $n=-4$. This value is somewhat
steeper than that discussed in Section~\ref{sec:star-counts}, but it
has the advantage of allowing the integrals to be solved analytically,
and hence leads to a simpler expression. In this case
Eq.~(\ref{eq:final_form}) becomes
\begin{equation}
\frac{\sigma_x^2}{\sigma_z^2} = \frac{1}{4 q^2(1 - q^2) {\rm arctg}
\biggl[\sqrt{\displaystyle\frac{1-q}{1+q}}\biggr]} 
\bigg\{q \sqrt{1-q^2} + 2 (1 - 2 q^2)
{\rm arctg}\biggl[\sqrt{\frac{1-q}{1+q}}\biggr]\bigg\}
\end{equation}

The observed flattening of $q \sim 0.6$ therefore translates into a
predicted $\sigma_x^2/\sigma_z^2 = 1/2 (\sigma_x^2 +
\sigma_y^2)/\sigma_z^2 = 1/2 (\sigma_R^2 + \sigma_\phi^2)/\sigma_z^2 =
1.73$.  The measured value of $1/2 ( \sigma_R^2 +
\sigma_\phi^2)/\sigma_z^2 = 1.76 \pm 0.37$ \citep{chiba-beers} is
interestingly close. This presumably also means that the effective
Galactic potential near the Sun is quite close to that of an
isothermal sphere \citep[see also][]{white85,sl-c}.

\section{Substructure}
\label{sec:subst}

One of the most fundamental ingredients of the $\Lambda$CDM
cosmological model is that galaxies grow via mergers. Therefore,
perhaps the most direct way of testing this paradigm is by quantifying
the amount of mergers that galaxies have experienced over their
lifetime. This implies finding the traces of those merger events.

\begin{figure}
\begin{center}
\includegraphics[width=0.6\textwidth]{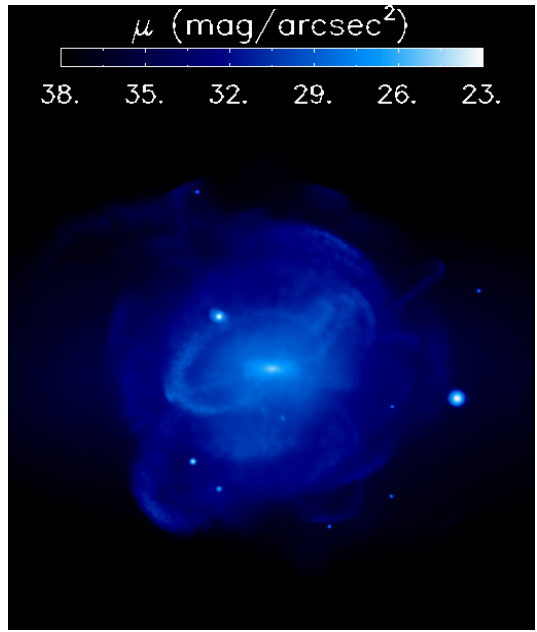}
\end{center}
\caption{A ``simulated'' stellar halo built up from accreted
satellites. The color scale indicates surface brightness. [Sanjib
Sharma, Kathryn Johnston and James Bullock are acknowledged for this
figure.]}
\label{fig:stellar-halo-acc}
\end{figure}

Mergers are expected to have left large amounts of debris in the
present-day components of galaxies. In particular, a stellar halo is
easily built up via the superposition of disrupted satellites
\citep{kvj96,hw99,bullock-01,paul}. This is clearly exemplified in
Figure \ref{fig:stellar-halo-acc} adapted from the simulations and
semi-analytic modeling by \cite{bj05}.

The prediction that debris in the halo should be ubiquitous in
hierarchical cosmologies, together with the discoveries of a
disrupting dwarf galaxy in the halo by \cite{ibata94} as well as of
nearby remains from an ancient accretion event by \cite{h99}, have
boosted the search for substructure in this component of our Galaxy in
recent years. Such searches have benefited enormously from the advent
of large photometric surveys, and the SDSS in particular, as we shall
see below, have all led to a shift in paradigm in the field over the
past 10 years.

\begin{figure}
\includegraphics[width=0.9\textwidth]{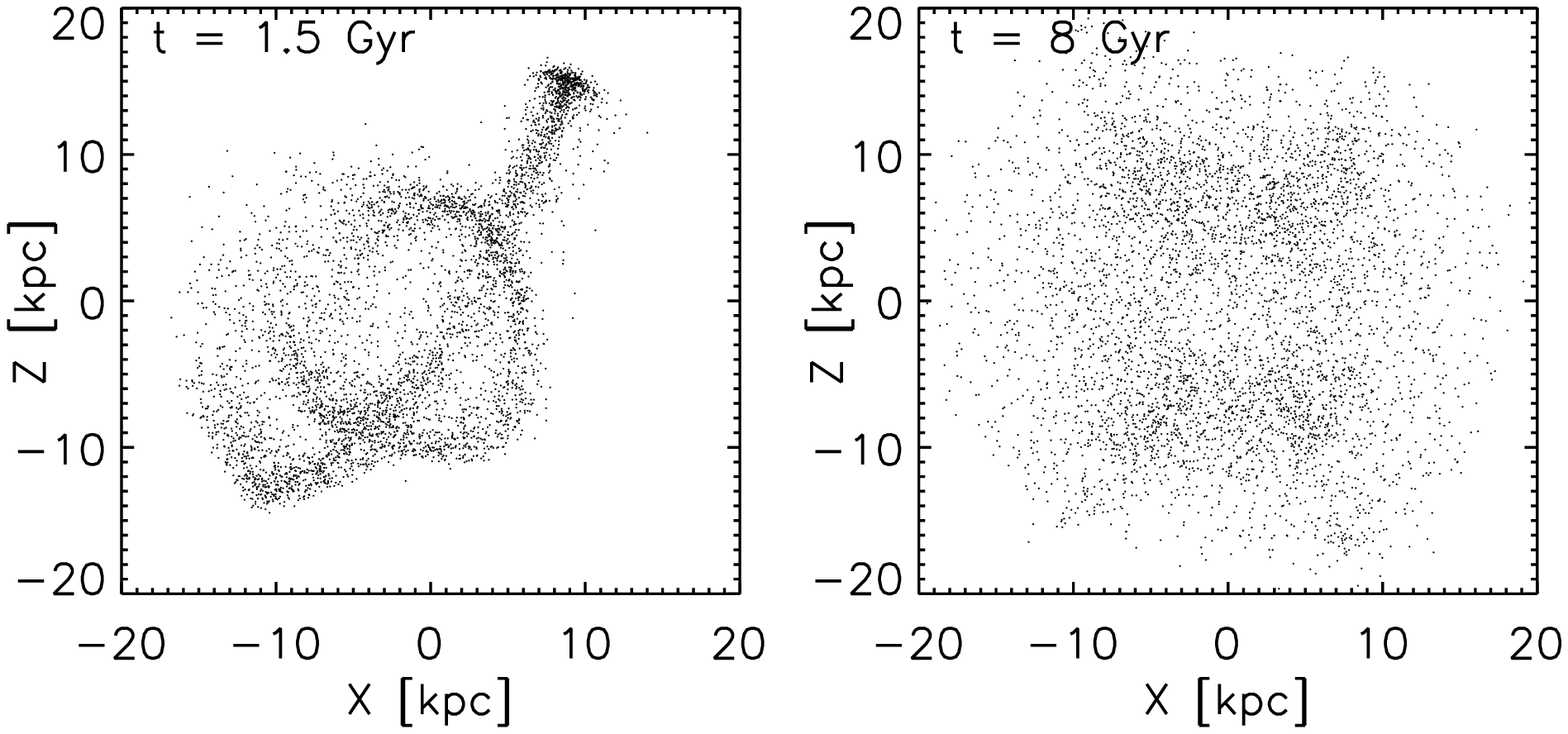}
\includegraphics[width=0.9\textwidth]{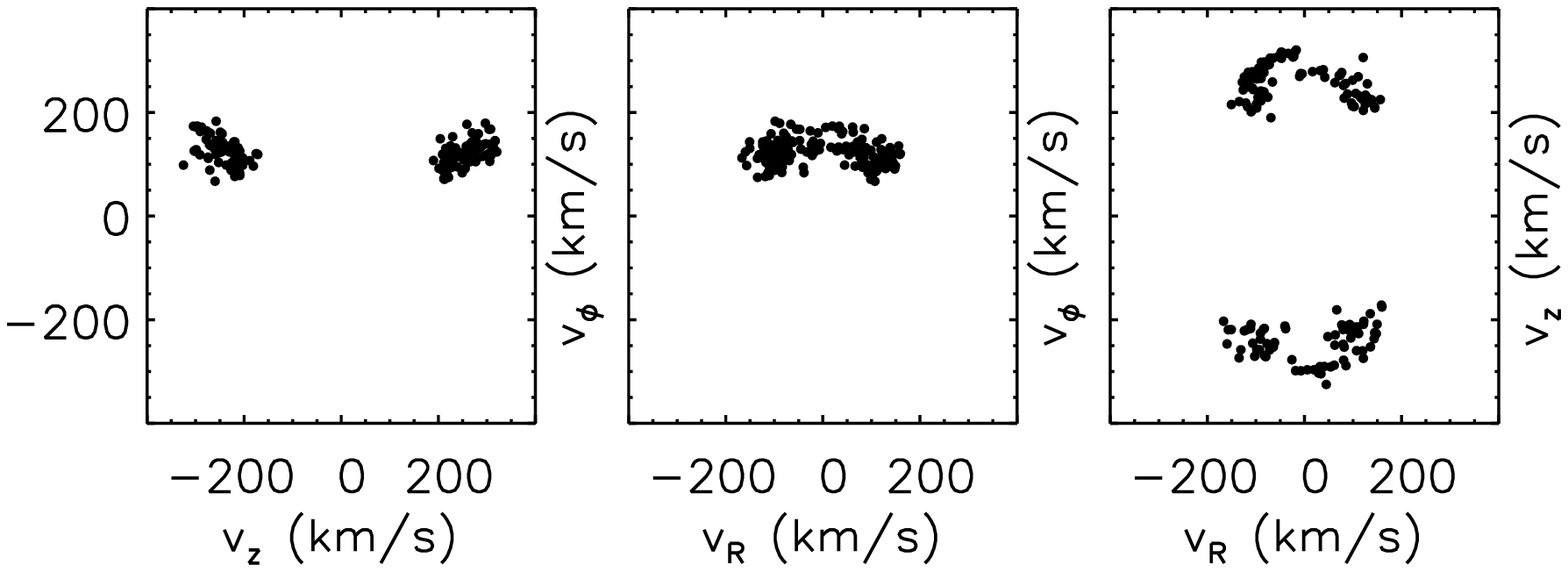}
\caption{Simulation of the disruption of a satellite galaxy orbiting
in the inner regions of the halo of the Milky Way. The top panels show
the $xz$ projection of the particles' positions, which evidence
significant evolution over short timescales. The velocities of the
particles located in a spherical volume of $\sim 2$ kpc around the Sun
at t$\sim 8$ Gyr are strongly clustered, as shown in the bottom
panels. Based on the simulation from \cite{h99}.}
\label{fig:h99}
\end{figure}

From a theoretical point of view, one expects substructures in the
spatial domain, as well as in velocity space (or in phase-space). One
can distinguish essentially two regimes in which substructures will be
most easily found in either of these domains:
\begin{itemize}
\item $t \sim t_{\rm orb}$ Short after infall, or for satellites orbiting
the outer halo (dynamical timescales are comparable to the time since
infall). In this case, streams are very coherent in space. Hence, a
simple way to reveal these accretion events is to map of the positions
of halo stars on the sky \citep{kvj96}.

\item $t >> t_{\rm orb}$ Long after infall, or in the inner halo ($r < 15 - 20$
kpc). The debris is spatially well-mixed, as shown in Figure
\ref{fig:h99}.  Because of the conservation of phase-space density,
the stars in a stream will have very similar velocities (see bottom
panels in Figure \ref{fig:h99}). In fact, streams get colder as time
goes by. Therefore the characteristic signature of an accreted galaxy,
even many Gyr after infall, is the presence of very cold structures in
velocity space \citep{hw99}.

\end{itemize}

\subsection{In the spatial domain}

Because of the relative ease with which young substructures may be
uncovered in the spatial domain, the largest fraction of known streams
are those that are spatially coherent.

These discoveries did not really happen systematically until the
advent of deep photometric wide-field surveys. The Sagittarius (Sgr)
dwarf galaxy was a true beacon, because it was clearly tidally
disturbed \citep{ibata94,gomez99}. Shortly after its discovery, a
southern extension was found by \cite{mateo98}. Somewhat later,
\cite{majewski99} reported debris at 40$\odeg$ further south from the
main body, which could be tentatively fit by models of a disrupting
dwarf \citep{kvj99}. A very large excess of RR~Lyrae and BHB stars at
high Galactic latitudes was subsequently discovered in the SDSS survey
\citep{ivezic,yanny00}. This overdensity was roughly along the
north-south elongation of the Sgr dSph, but now at 60$\odeg$ away from
it, and at a significantly larger distance (50 kpc instead of 15 kpc
from the Galactic center).  The association of the overdensity to the
Sgr dwarf galaxy northern/leading stream was nearly obvious \citep[see
e.g.][]{gomez99}, as the models of \cite{hw01} showed\footnote{This
was posted as a preprint on astro-ph in March 2000; just as the SDSS
discoveries were being made.}.

After this, the Spaghetti Survey also found an overdensity at the same
location using red giant branch stars with measured radial velocities
\citep{robbie}. \cite{martinez-delgado01} using deep color-magnitude
diagrams detected the presence of a main sequence and turn-off at
similar locations, and \cite{ibata01} using carbon stars were able to
trace the stream over an extended portion of the sky.

\begin{figure}
\begin{center}
\includegraphics[width=0.9\textwidth]{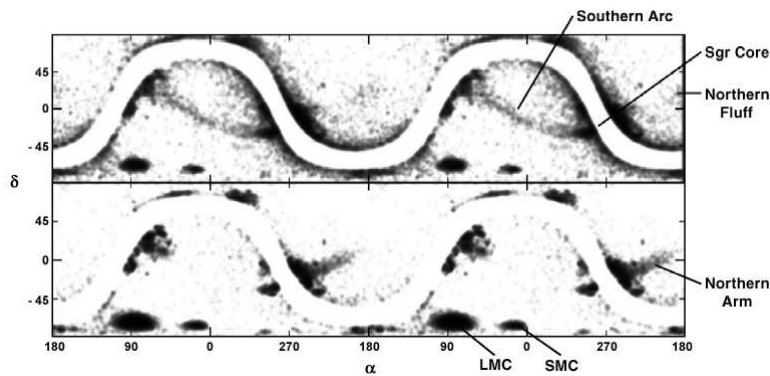}
\end{center}
\caption{2MASS revealing the streams from Sgr. Smoothed maps of the
sky for point sources selected according to $11 \le K_s \le 12$ and
$1.00 < J-K_s < 1.05$ (top), and $12 \le K_s \le 13$ and $1.05 < J-K_s
< 1.15$ (bottom). Two cycles around the sky are shown to demonstrate
the continuity of features. [From \cite{Majewski03}. Courtesy of Steve
Majewski. Reproduced with permission of the AAS.]}
\label{fig:sgr2mass}
\end{figure}

A spectacular view of the Sgr streams was given by the Two Micron All
Sky Survey (2MASS\footnote{\sf
http://pegasus.phast.umass.edu/}). \cite{Majewski03} using M giants
(selected from the 2MASS database on the basis of their colors) were
able to map the streams 360$\odeg$ around the sky as shown in Figure
\ref{fig:sgr2mass}. The vast extent of the Sgr dwarf streams has been
used to constrain the shape of the Galactic gravitational potential,
albeit with conflicting results \citep{helmi04,kvj05,fellhauer06}.

\begin{figure}
\includegraphics[width=0.9\textwidth]{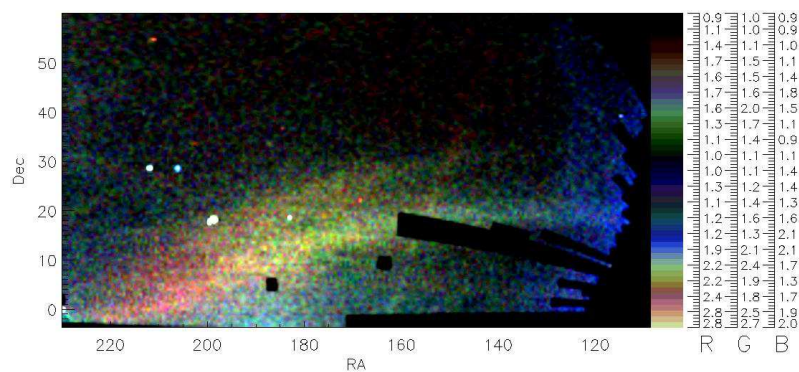}
\caption{The field of streams from SDSS featuring Sgr, several new
satellites and the low-latitude Monoceros ring. [From
\cite{vasily-fos}. Courtesy of Vasily Belokurov. Reproduced with
permission of Wiley-Blackwell Publishing.]}
\label{fig:fos}
\end{figure}

In the years to follow, many of the structures discovered in the halo
would be linked to the Sgr debris
\citep{newberg,vivas-sgr,martinez-delgado04}. An exception to this is
the Monoceros ring \citep{yanny03,ibata}, a low-latitude stream
towards the anticenter of the Galaxy, that spans about 100$\odeg$ in
longitude at a nearly constant distance. It is somewhat of a semantic
distinction to state that this structure is part of the stellar halo,
or of the Galactic thin or thick disks.  As yet, the nature of the
feature and of the Canis Major overdensity in a similar location at
low Galactic latitude \citep{martin04}, are highly debated.  Possible
interpretations are that it is debris from an accreted satellite
\citep[e.g.][]{helmi03,martin04,penarrubia05,martinez-delgado05}, an
overdensity associated to the Galactic warp \citep{momany04,momany06},
or even a projection effect caused by looking along the nearby
Norma-Cygnus spiral arm \citep{moitinho}. Recently, \cite{grillmair}
has mapped the structure of the Monoceros overdensity using SDSS, and
showed that it consists, at least in part, of a set of very narrow
low-latitude substructures (see Figure \ref{fig:fos}), which can
easily be explained by the models of \cite{penarrubia05}. It is
worthwhile mentioning here that such a satellite may well perturb the
Galactic disk, and induce departures from axial symmetry or
overdensities such as e.g. Canis Major. Furthermore, the Sgr dwarf
leading tail also appears to be crossing the Galactic plane at roughly
the same location and distance as the ring \citep{newberg07}, and may
well be responsible for some of the substructures observed.

At the time of writing of this review, other substantial substructures
have been uncovered, such as the Virgo overdensity \citep{juric},
possibly linked to the Virgo Stellar Structure (\cite{vivas-sgr};
confirmed by \cite{duffau} using radial velocities), the
Triangulum-Andromeda \citep{rocha-pinto} and the Hercules-Aquila
clouds \citep{vasily-ha}. The nature of these substructures is
unclear, particularly because of the large area on the sky that they
cover. It is evident that detailed kinematic studies of large samples of
stars with good distances, as well as chemical abundances, and full
coverage of the structures, are necessary to understand whether these
are truly related to tidal debris. In this case, one would expect to
see coherent motions, as indeed appears to be the case for the
Hercules-Aquila cloud. 

Narrow streams have been detected around globular clusters
\citep{grillmair95}, the most spectacular examples being Pal 5
\citep{oden} and NGC5466 \citep{grillmair-ngc,vasily-ngc}. This shows
that globular clusters probably also have contributed to the halo
field population \citep{parmentier}. Another example of a narrow tidal
tail is the Orphan-Stream \citep{grillmair-ostream,o-stream}, whose
progenitor may well have been a dwarf spheroidal-like object accreted
by the Galaxy several Gyr ago \citep{sales08}.

It is very important to understand what kind of biases affect our
ability to detect substructure, particularly if our goal is to
determine the mass function of the objects accreted and the timing of
these events. For example, it is obviously easier to detect high
surface brightness features like those originating in massive recently
accreted satellites. For instance, the Sagittarius stream reaches a
surface brightness of $\Sigma_V \sim 28$ mag/arcsec$^2$ at $l \sim
350\odeg$ and $b \sim 50\odeg$ \citep{martinez-delgado01}, where it
was first detected, while the other overdensities discovered so far
typically have surface brightnesses $\Sigma_V \sim 32-33$
mag/arcsec$^2$ \citep[see Figure 8 in][and private
communication]{vasily}.

The large majority of the above-mentioned substructures are located in
the outer halo, at distances $\simmore 15$ kpc from the Galactic
center.  The inner halo, on the other hand, appears to be relatively
smooth \citep{lemon}.

\subsection{In the velocity domain}

As briefly discussed in Sec.~\ref{sec:kinematics}, kinematic
information is more expensive to obtain, and hence the velocity domain
has been less explored despite its significantly larger suitability
for searches of substructure.

Two new major spectroscopic surveys are likely to change this in the
nearby future. The Sloan Extension for Galactic Understanding and
Exploration \citep[SEGUE\footnote{\sf
http://www.sdss.org/segue/aboutsegue.html},][]{beers-segue} and the
RAdial Velocity Experiment \citep[RAVE\footnote{\sf
http://www.rave-survey.aip.de/rave/},][]{steinmetz} are both currently
collecting spectra for very considerable samples of stars. The final
catalogs will be highly complementary. SEGUE will observe $\sim
240,000$ stars in the range $15 < V < 21$, while RAVE aims at 10$^6$
stars with $9 < I < 12$. The average velocity errors that these
surveys can achieve are of the order of 10\kms and 1\kms
respectively. The advent of such large spectroscopic surveys is
clearly a very interesting development since the kinematics of halo
stars can be used to find and constrain the nature of the
overdensities, as well as to measure the mass distribution in our
Galaxy \citep[e.g.][]{smith-rave}.

The catalogs of halo stars with 3D kinematics used so far in searches
of substructure are significantly smaller and less accurate \citep[see
for example, the work
of][]{chiba-yoshii,h99,chiba-beers,kepley}. Thanks to the developments
mentioned above, however, we may expect soon a ten-fold increase in
the number of nearby halo stars with full kinematics. For example it
will be possible to combine the RAVE radial velocities with the
relatively accurate proper motions from the Tycho-2
catalog. (\cite{carollo} have already demonstrated the potential of
combining the SDSS data (with radial velocities and stellar
parameters) with proper motions from the USNO-B catalog, to
understand the global structure of the halo, see
Section~\ref{sec:mdf}. Note that this dataset is not optimal to
search for substructure in the Solar neighborhood because of the
relatively large velocity errors, which are typically in the range
30--50\kms).

So far, there have only been a few detections of substructure near the
Sun. This is not really surprising. Because of the short dynamical
timescales in the inner Galaxy (a typical halo star near the Sun will
have made between 50 and 100 revolutions in a Hubble time), each
accreted satellite will generally give rise to multiple streams (or
wraps of debris) that will cross the Solar neighborhood, as shown in
Figure~\ref{fig:h99}. In a stellar halo completely built from
disrupted galaxies, the predicted total number of nearby streams is
\begin{equation}
N_{\rm streams} \sim \{300-500\} (t/10 \,{\rm Gyr})^3
\end{equation}
i.e., a few hundred streams are expected to cross the solar
neighborhood after a Hubble time \citep{hw99}. This quantity is only
weakly dependent on the number and mass spectrum of the accreted
satellites. Furthermore, in the context of the $\Lambda$CDM model, it
is worthwhile stressing that these streams are expected to originate
in only a handful of objects \citep{hws03}.

To detect these streams, samples with approximately 5000 halo stars
are needed (that is, $N_{\rm sample} \sim 10 \times N_{\rm
streams}$). Given the size of the samples currently available, we
could clearly be in the limit in which essentially each stream is
populated by just one star.

Nevertheless, \cite{h99} successfully detected two streams coming from
a common satellite accreted $6 - 9$ Gyr ago \citep{kepley}. A
simulation to reproduce its properties is shown in Figure
\ref{fig:h99}. The reason that these authors were able to detect the
streams is three-fold. Firstly, they studied the distribution of stars
in an ``integrals of motion'' space (defined by two components of the
angular momentum), where stars from a common progenitor are very
strongly clustered \citep{hdz}. Thus they could map the two streams,
one with 4 stars and one with 8 stars, to one overdensity with 12
stars.  The second reason for the successful detection is that the
stream stars are going through their pericenters, and at the orbital
turning points one expects an enhancement in the density of a
stream. \citep[This phenomenon is in essence the same as the shells
observed around elliptical galaxies,][]{quinn}. This explains why the
streams detected by \cite{h99} were not really that cold. The third
reason is that the values of the integrals of motion of the accreted
stars are quite different from those typical of solar neighborhood
halo stars --lying in the tails of the distribution, exactly where one
expects the most recently accreted objects to be.

\cite{seabroke} have recently put constraints on the presence of
nearby massive halo streams using the radial velocities of stars
towards the Galactic poles. They find no vertical flows with densities
greater than 1.4 stars/kpc$^3$. This rules out the possibility that
streams from the Sagittarius dwarf are crossing the Solar
neighborhood, which in turn implies that models in which the dark
matter halo is prolate are favored \citep{helmi04,law}.

Seabroke's results are in excellent agreement with the arguments of
\cite{gould}, who constrained the degree of granularity using the
revised New Luyten Two-Tenths (rNLTT) catalog of proper motions of
4500 nearby halo stars. He finds, using statistical arguments, that no
streams crossing the Sun contain more than 5\% of the nearby halo
stars. This is also in unison with the results of Helmi et al. (1999).

As mentioned above, further detections of substructure will require
significantly larger samples with full and accurate 3D kinematics. The
first steps towards this will come with RAVE and SEGUE, but the clear
breakthrough will be made by the European Space Agency's mission Gaia
\citep{perryman}. Gaia is scheduled for launch in 2011 and will
provide high accuracy astrometry (at the 10 microarcsecond level for
$V \le 12$ stars, and 200 microarcseconds for $V \sim 20$) and
multi-color photometry for over 1 billion stars throughout the Milky
Way. Radial velocities at 10\kms accuracy will be acquired for stars
down to $V \sim 16-17$.  Therefore, the Gaia dataset will allow us to
quantify the amount of substructure in the halo of our Galaxy, and
consequently its evolutionary path.

\subsection{Implications}

Much of the work in the field of ``substructure'' has so far been
rather qualitative in nature.  The discoveries of the substructures
are clearly the first step, but what do they actually tell us about
how our Galaxy formed?  This requires a more quantitative approach.

\begin{figure}
\includegraphics[width=0.9\textwidth]{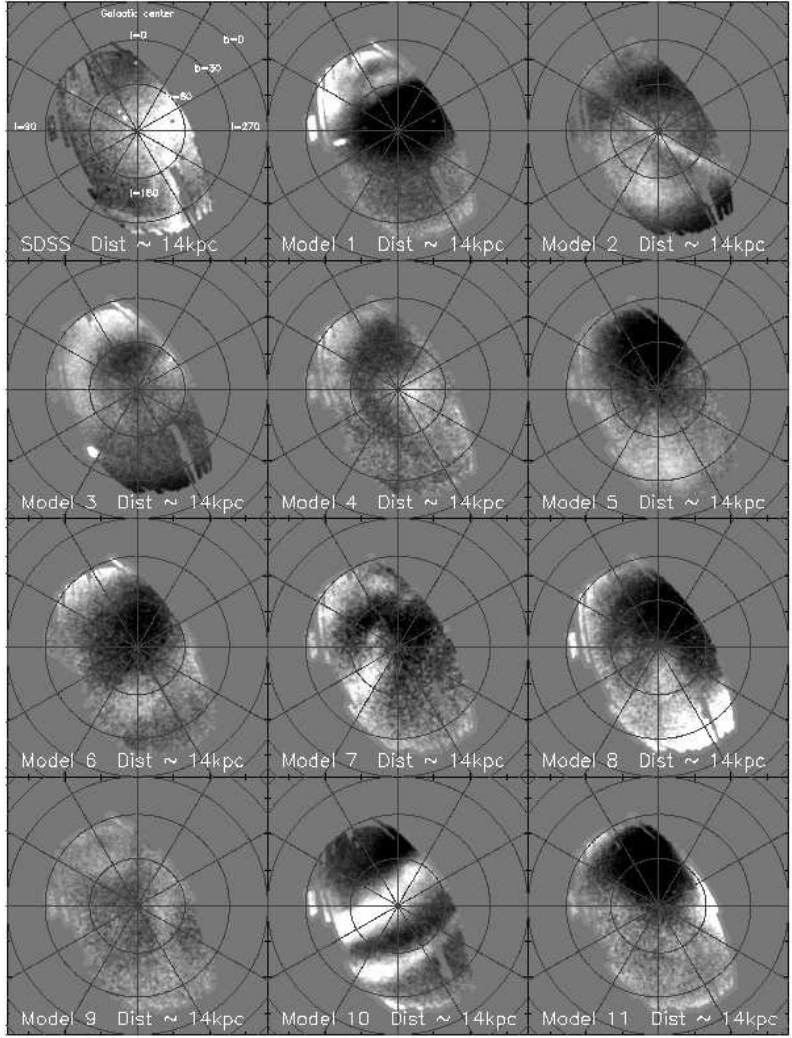}
\caption{Residuals (SDSS or simulations minus the smooth model)
smoothed using a $\sigma = 42'$ Gaussian from the best {\it oblate}
model fits for the SDSS data (top left panel) and for the 11
simulations from \citet{bj05}.  Only the $20 \le r < 20.5$ slice,
corresponding to heliocentric distances $\sim 14$\,kpc are
shown. [From \cite{Bell}. Courtesy of Eric Bell].}
\label{fig:bell}
\end{figure}

\cite{Bell} have used SDSS data to compute the {\it rms} fluctuations
in the number of objects compared to a smooth halo model. They find
typical values of $\sim 40\%$. They interpret this using the
simulations of \cite{bj05}, where this statistic can vary from 20\% to
80\% depending on the detailed halo formation history.  This could
imply that the (outer) Galactic halo is fairly average in its history
and could have been built completely by accreted satellites.  However,
a careful comparison of the overall structure of the fluctuations
shows that there may still be room for a smooth halo component, as
illustrated in Figure \ref{fig:bell}. For example, there are
significant under-densities in the \cite{bj05} maps, which are not
apparent to the same extent in the SDSS data.

\cite{else} use a different approach and data from the Spaghetti
Survey to quantify the amount of substructure. Their sample contains
approximately 100 red giant stars with radial velocity and distance
information. They develop a statistic called the 4-distance, which is
essentially a correlation function in a 4D space:
\begin{equation}
4d_{ij} = \sqrt{a_\omega(\omega_{ij})^2 + a_d(d_i - d_j)^2 +
a_v(v_{{\rm rad},i} - v_{{\rm rad},j})^2}
\end{equation}
where $\omega_{ij}$ denotes the angular separation on the sky of star
$i$ with respect to star $j$, $d_i$ its distance and $v_{{\rm rad},i}$
its radial (line-of-sight) velocity. The coefficients $a_x$ denote
weights to normalize the various observables and to account for the
observational errors. Substructure would manifest itself in an excess
of nearby pairs in this 4 dimensional space. Using this method,
\cite{else} find that $\sim 20$\% of the stars are in substructures,
the most prominent of which corresponds to the Sgr stream northern
overdensity \citep{robbie}. When comparing to the simulations of halos
built up completely from accreted satellites of $10^7 \sm$ by
\cite{paul}, \cite{else} find that these typically show a
significantly larger amount of substructure. Only in about 10\% of the
cases, the characteristics of such halos are similar to what is
measured by the Spaghetti survey. They interpret this as meaning that
if the whole halo was built from accreted satellites, these must have
left behind relatively broad streams, implying that either they were
relatively large or have orbited the Galaxy long enough to be
significantly broadened. The other possible interpretation is that
there is an underlying smooth component in the halo. \cite{else} were
clearly limited by the small sample size, which provided significant
Poisson noise to their statistical estimations. This situation is
likely to change in the very near future with SEGUE, which will
provide orders of magnitude larger samples of red giant stars with
accurate distances and velocities \citep{morrison08}.

\section{Ages and chemical abundances of halo stars}
\label{sec:chemistry}

\subsection{Age distribution}

The determination of the age of individual stars is one of the most
difficult measurements in astrophysics. Some possible techniques
include isochrone fitting of individual stars (their absolute
magnitude, effective temperature and metallicity needs to be
accurately known) or star clusters; asteroseismology and
nucleochronology.

Direct age measurements can be inferred by comparing the observed
abundance ratios of the radioactive elements of e.g. $^{232}$Th
(half-life of 14 Gyr) and $^{238}$U (4.5 Gyr) with theoretical
predictions of their initial production. The first measurements of
very metal-poor nearby stars have yielded ages of $15.2 \pm 3.7$ Gyr
\citep{sneden}, 14 $\pm$ 3 Gyr \citep{cayrel01,hill02}, and 13.2 Gyr
\citep{frebel}.

The white dwarf luminosity function, and especially its cut-off, also
provides strong constraints on the age of the population to which they
belong \citep[see, e.g.][]{chaboyer98}. This method, based on the
white dwarf cooling sequence, has been used in the past to derive the
age of the oldest stars in the Galactic disk \citep[see
e.g.][]{hansen}. However, the scarcity of known halo white dwarfs and
the lack of good kinematical data to distinguish these from the far
more common disk stars have so far, prevented such a measurement for
the halo \citep{torres}.

The horizontal branch morphology of a stellar population at a given
abundance tends to be bluer for older populations, and redder for
younger ones \citep{layden98b}. In the halo, there is some indication
that the horizontal branch morphology of field stars becomes redder
with Galactocentric distance. If this is interpreted as due to age, it
means that the outer halo would be younger by a few Gyr, compared to
the inner halo \citep[see e.g.][]{sz,suntzeff}. It should be noted,
however, that the existence of an age spread or a gradient in the
population of halo globular clusters is still under debate
\citep[e.g.][]{sarajedini,stetson-gc,de-angeli}.

\subsection{Chemical abundances}
\subsubsection{Metallicity distribution}
\label{sec:mdf}

The local halo metallicity distribution peaks at a value $\feh \sim
-1.6$~dex, and extends well below $\feh\sim -3$ dex
\citep[e.g.][]{ryan-norris}.  

\begin{figure}
\includegraphics[width=0.9\textwidth,angle=90]{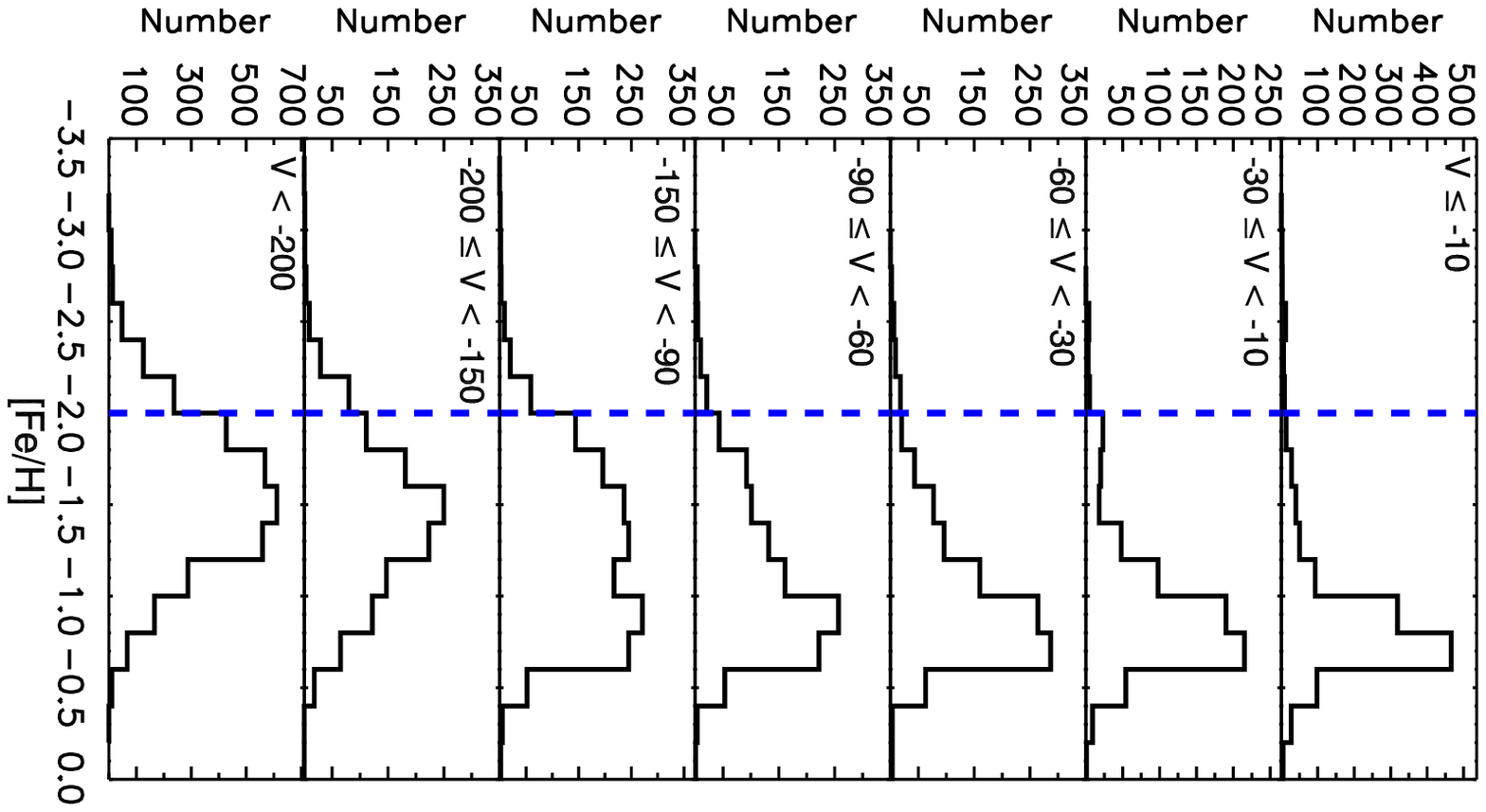}
\includegraphics[width=0.9\textwidth,angle=90]{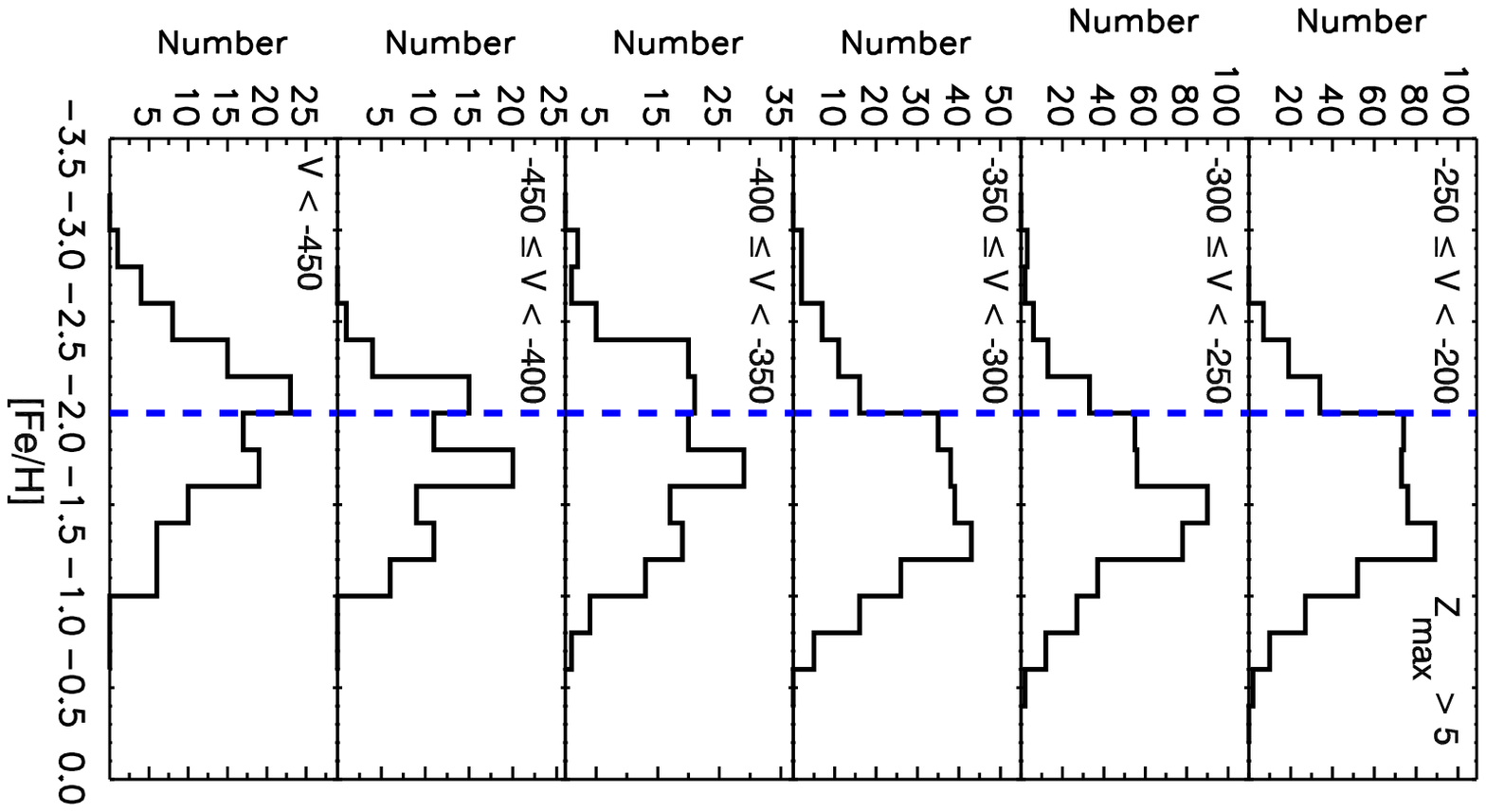}
\caption{Metallicity distribution function based on medium-resolution
spectroscopy for $2 \times 10^4$ nearby stars obtained with SDSS. The
histograms show that the inner halo ($V \sim -200$ \kms) exhibits a
peak metallicity $\feh = -1.6$, and that the outer halo (right panels)
peak at a value $\feh = -2.2$, and have a net retrograde
rotation. [From \cite{carollo}. Courtesy of Daniela
Carollo. Reproduced with permission of Nature.]}
\label{fig:carollo}
\end{figure}

Recently \cite{carollo} based on SDSS data revived the idea that the
halo may be described by two broadly overlapping components. The inner
halo peaks at $\feh \sim -1.6$ dex, is flattened and slightly
prograde, while the outer halo would peak around $\feh \sim -2.2$ dex,
would be rounder and in net retrograde rotation. Although a dichotomy
in the Galactic halo had been suggested before (see also
Section~\ref{sec:star-counts}), this has now been confirmed with a much
larger sample of stars covering a large area on the sky and is now
clearly reflected also in the metallicity distribution as shown in
Figure~\ref{fig:carollo}.

Because of the importance of metal-poor stars in constraining the
formation history of the Galaxy, as well as the physics of the
high-redshift Universe (e.g. reionization, initial mass function, etc)
significant efforts have been made to better characterize the
metallicity distribution at the very metal-poor end. More recently,
the HK and the Hamburg/ESO (HES) surveys have yielded a dramatic
increase in the number of nearby metal-poor stars known
\citep{beers-christlieb}.  In these surveys, metal-poor candidates are
selected from objective-prism spectra for which the Ca~H and K lines
are weaker than expected at a given $(B-V)$ color. These stars are
then followed up with medium-resolution spectroscopy \citep[see, for
example,][]{frebel06}.

These surveys have demonstrated two important points. First, three
extremely metal-poor stars with $\feh < -4.75$ dex have been
discovered, all of which are carbon rich. Secondly, no metal-free
stars (Population III) have been found thus far.

These results could in principle, put strong constraints on the
initial mass function (IMF). For example, \cite{ssf} find, using
semi-analytic models of galaxy formation, that the lack of metal-free
stars constrains the critical metallicity for low-mass star formation
to be $Z_{\rm cr} > 0$, \citep[see][]{schneider}, or the masses of the
first stars to be $M_{\rm PopIII} > 0.9 \sm$. \cite{brook07} reach a
similar conclusion using cosmological hydrodynamical simulations.

From a theoretical point of view, the initial mass function of
Population III stars is expected to be top-heavy, with masses in the
range $100-1000 \sm$, e.g. \cite{bromm99,abel}. Such stars are
predicted to leave very characteristic imprints in the abundance
patterns of the next generation of stars (a very strong odd-even
effect, e.g. \cite{heger}). However, such patterns are not consistent
with those observed for example by \cite{cayrel04} for nearby very
metal-poor stars. Furthermore, the observed trends can be explained by
normal supernovae with masses $< 100\sm$ \citep{umeda,silk06}.

The solution to this conundrum is still to be found. Given the small
number statistics, the IMF appears to be consistent with Salpeter as
well as with a top-heavy functional form \citep{ssf}. On the other
hand, the high incidence of carbon-enhanced, unevolved stars among
extremely metal-poor stars has been used to propose that the IMF in
the early Galaxy was shifted toward higher masses
\citep{lucatello}. Based on the idea that the two most metal-poor
stars known could be both in binary systems (and that their high
[C/Fe] could be due to the winds from an AGB companion star),
\cite{tumlinson} discusses that an IMF with a characteristic mass
$\sim 3 \sm$ would reproduce the properties and distribution of the
halo stars at the metal-poor end.

In the coming years, our knowledge of the halo metallicity
distribution function is likely to improve significantly with surveys
such as Skymapper\footnote{\sf
http://www.mso.anu.edu.au/skymapper/}, SEGUE and
LAMOST\footnote{\sf http://www.lamost.org}, which will extend by large
factors the currently known handful of extremely metal-poor stars
\citep{christlieb06}. These new datasets will give us to access to the
very high-redshift Universe through the fossil record left by the
first generations of stars ever formed.

\subsubsection{Elemental abundances}
\label{sec:elements}

Much more information about the chemical history of the Galactic halo
may be obtained from detailed studies of the elemental abundances
patterns. This is because the various chemical elements are
synthesized by different processes in stars of different masses and on
different timescales \citep{mcwilliam}.  For example,
$\alpha$-elements (O, Mg, Si, Ca, S, and Ti) are produced during the
explosion of a massive star as a Supernova Type II, which occurs only
a few million years after its formation. Iron-peak elements, on the
other hand, are produced both by Type Ia and Type II supernovae. Type
Ia supernovae are the result of a thermonuclear explosion induced by
the transfer of mass onto the surface of a white dwarf by a binary
companion star. This implies that these explosions take place
typically on a longer timescale, of the order of $0.1$ to a few Gyr
\citep[e.g.][]{matteucci-typeI}. Heavier elements beyond the iron-peak
are created by neutron capture, through slow (s) and rapid (r)
processes. The s-process can take place when the neutron flux is
relatively low, i.e. the timescale between neutron captures is large
compared to that of the beta decay. These conditions are found in the
envelopes of Asymptotic Giant Branch (AGB) stars, and are most
efficient in stars with masses $3 - 5 \sm$, implying a timescale of
100 Myr after the stars were born \citep[see][and references
therein]{goriely}.  On the other hand, the r-process occurs when the
neutron flux is sufficiently high to allow for rapid neutron
captures. This is believed to occur in environments like those
produced by supernovae Type II. It should be stressed, however, that
the exact sites and conditions under which the various neutron-capture
elements are produced, particularly at very low metallicities, are
still somewhat uncertain \citep{travaglio,arnould}.

\begin{figure}
\begin{center}
\includegraphics[width=0.5\textwidth,angle=270]{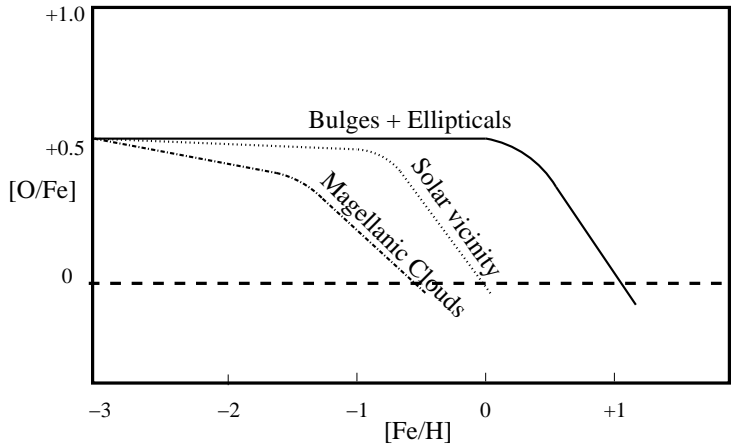}
\end{center}
\caption{Sketch of the predicted [O/Fe] versus [Fe/H] relation in
different systems as a consequence of their different enrichment
timescales.  The $\alpha$ enhancement seen in the Galactic bulge can
be explained by an intense star formation burst in which mainly
supernovae Type II polluted the interstellar medium. On the other
hand, the Magellanic Clouds had a lower star formation rate, and so
most of their stars have a relatively enhanced iron-content. Thin-disk
stars near the Sun are an intermediate case. [Based on
\cite{matteucci}.]}
\label{fig:chem-evol}
\end{figure}

Figure~\ref{fig:chem-evol} shows how the chemical enrichment depicted
by the [$\alpha$/Fe] versus $\feh$ may proceed in a given object
depending on its initial mass function (which sets the initial level
of [$\alpha$/Fe]) and its star formation rate (which sets at what
metallicity the ``knee'' is observed).

Interestingly, the majority of the local halo stars are
$\alpha$-enhanced \citep{wheeler,nissen94,carretta}, where typically
[$\alpha$/Fe] $\sim 0.3$ dex. This implies that the formation
timescale for these stars was very short, less than 1 Gyr. This
argument is based on the fact that the $\alpha$-enhancement may also
be seen as an iron deficiency. Such a configuration may be obtained
when the interstellar medium (ISM) has been enriched by supernovae
Type II, but not significantly by Type Ia \citep{gw98}.

\begin{figure}
\includegraphics[width=0.5\textwidth]{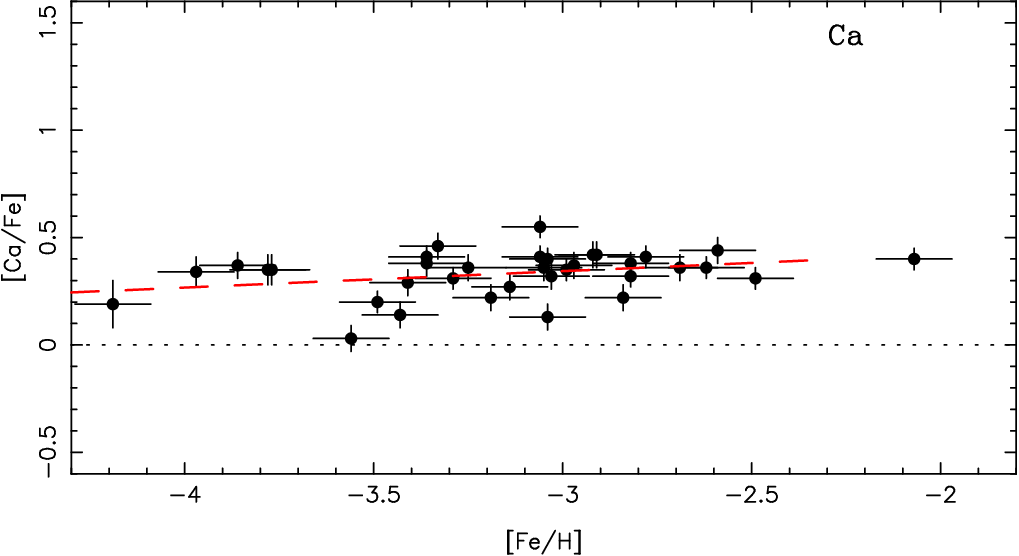}
\includegraphics[width=0.5\textwidth]{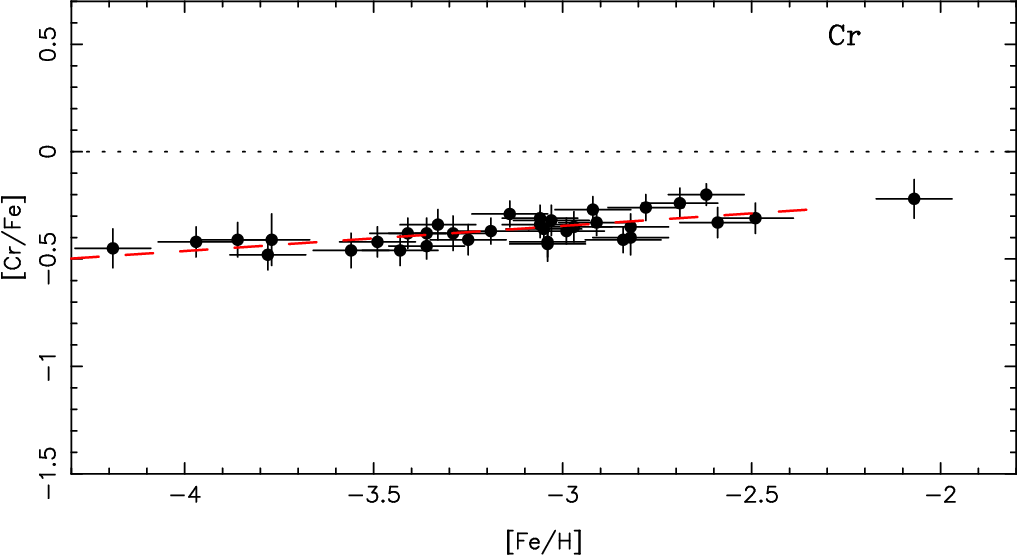}
\caption{Elements ratios as function of $\feh$ for the sample of very
metal-poor stars from \cite{cayrel04}. Ca is an $\alpha$ element,
while Cr is from the iron-peak, yet in both cases the scatter is
extremely small. In contrast, neutron-capture element ratios such as
[Y/Fe] and [Ba/Fe] show very large scatter in the same stars
\citep{francois}.[From \cite{cayrel04}. Courtesy of Rogier
Cayrel. Reproduced with permission of A\&A.]}
\label{fig:cayrel}
\end{figure}

More detailed insights have been provided by \cite{cayrel04} and
\cite{francois}. They have obtained high-resolution spectra ($R \sim
45000$) for 30 very metal-poor giants ($-4.1 < \feh < -2.7$
dex). \cite{cayrel04} find that the scatter in several [$\alpha$/Fe]
and [Fe-peak/Fe] ratios is consistent with being only due to
observational errors, as shown in Figure~\ref{fig:cayrel} \citep[see
also][]{cohen}. This would suggest that these stars formed in an
extremely well-mixed medium, i.e. the cosmic scatter expected from the
pollution by supernovae of different masses is not observed. In
contrast, the neutron-capture elements behave very differently,
showing a large spread at low metallicity \citep[][and references
therein]{barklem,francois}. It is currently unclear how to reconcile
these results. In a scenario in which the halo was formed via the
accretion of several independent objects, one would naively expect
that each of these objects followed their own chemical evolution path,
hence a superposition of their debris should give rise to a large
scatter in the various chemical abundance patterns. Perhaps the
observed large scatter in [neutron-capture/Fe] would be consistent
with such a scenario \citep[see][]{cescutti}, but why is there not a
spread in other elements?  Uncertainties in how nucleosynthesis
proceeds at very low metallicities, as well as in modeling the
chemical evolution in the proto-galaxy in a cosmological context, are
likely to be both elements of this puzzle.

The chemistry of stars can also be used to trace the merging history
of the Galactic halo \citep{freeman}. For example, \cite{nissen97}
found a correlation for local halo stars such that those with large
apocentric distances have relatively low [$\alpha$/Fe]. This may be an
indication that these anomalous halo stars have been accreted from
dwarf galaxies with a different chemical evolution history \citep[see
also][]{fulb,ivans}. This is consistent with the fact that stars in
dwarf spheroidal galaxies, tend to have lower [$\alpha$/Fe] at a given
metallicity compared to the halo. This in turn has been used to show
that dwarf galaxies could not have been significant contributors of
the stellar halo if accreted after they formed a significant fraction
of their stars \citep{shetrone,tolstoy03,venn04}.

\begin{figure}
\includegraphics[width=1.0\textwidth]{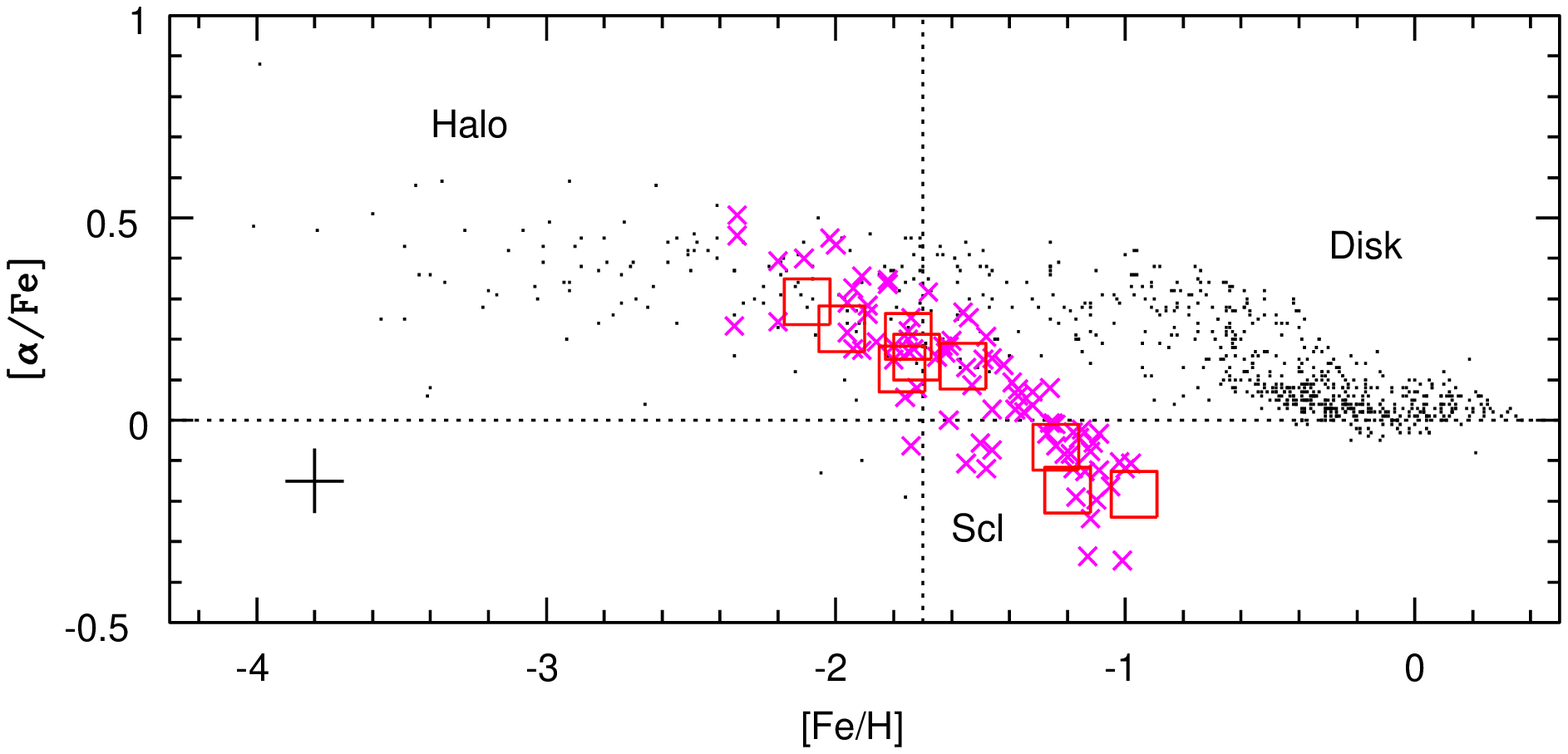}
\caption{$\alpha$-abundance (average of Ca, Mg and Ti) as function of
[Fe/H] for 92 member stars in a central field of Sculptor
\citep[crosses, from][]{hill08} compared to those in our Galaxy
\citep[dots; from the compilation of][]{venn04}. The open squares are
previous measurements of individual stars \citep{shetrone03}. The most
metal-poor stars in the dwarf spheroidals are as $\alpha$ enhanced as
field stars in the Galactic halo. It is only at higher $\feh$ that the
[$\alpha$/Fe] is significantly lower in the dwarf spheroidals
\citep{tolstoy06}. [Courtesy of Eline Tolstoy and Vanessa Hill.]}
\label{fig:bruno}
\end{figure}

More recent abundance studies of large samples of stars in several
dwarf spheroidals have demonstrated that these objects have similar
initial levels of [$\alpha$/Fe] at very low $\feh$, as shown in Figure
\ref{fig:bruno} \citep{tolstoy06,letarte,hill08}. It is likely that
their relatively low star-formation rates have led to slow chemical
enrichment, implying low values of [$\alpha$/Fe] compared to the halo
for all but the oldest (or most metal-poor) stars in the dwarf
spheroidals \citep{tolstoy03}.

\section{The satellites: dwarf galaxies and globular clusters}
\label{sec:satellites}

Approximately 150 globular clusters and about 20 satellite galaxies
are found in the Galactic halo. It is interesting to study their
distribution as well as their internal properties to elucidate the
evolutionary path of the Galaxy. In particular, it is worthwhile
comparing these objects' stellar populations to the halo field stars,
since in the most naive interpretation of the hierarchical paradigm,
the stars have their origin in small galaxies, presumably akin to a
certain (to be defined) extent the dwarf satellites we see around us
today.

The number of known dwarf galaxies in the Local Group has doubled in
the past few years, particularly at the faint luminosity end, thanks
to the discovery of a large number of very low surface brightness
satellites \citep[e.g.][]{vasily,martin07}. Despite this recent large
increase in the abundance of Galactic satellites, this is still
well-below the number of bound substructures orbiting galaxy-size
halos in pure cold dark matter simulations \citep{koposov}. This
discrepancy has been termed the ``missing satellites problem''
\citep{klypin99,moore99}. Several solutions have been put forward
including, for example, a modification of the nature of dark matter
\citep{spergel}. On the other hand, the modeling of the baryonic
physics at the low-mass end of the galaxy scale is far from complete,
and this is where environmental effects and processes such as
reionization, feedback from massive stars, etc.\ are likely to have a
significant impact on the probability of a substructure to hold onto
its gas and form stars \citep[e.g.][]{maclow,bullock-00,kravtsov04}.

The newly discovered satellites have luminosities in between (although
sometimes comparable to) globular clusters and the classic dwarf
spheroidals, but they are significantly more extended than the
former. They have extremely large mass-to-light ratios
\citep{simon-geha}, computed under the assumption of virial
equilibrium (which may very well turn out to be a very poor
approximation since these objects could well be tidal debris).

There are two very interesting properties of the spatial distribution
of the satellites. Firstly, their distribution is rather centrally
concentrated, in the sense that the median distance is less than 30\%
- 40\% of the virial radius \citep[as computed by][]{gb05,gb05err}. For
comparison, satellite substructures around halos in dark matter
cosmological simulations tend to be much more evenly distributed
\citep{gao,kang}. This apparent discrepancy can be solved by assuming,
for example, that the luminous satellites are embedded in those
substructures that were most massive at the time they crossed the
virial radius of the Galaxy. Also the radial distribution of the halo
globular clusters follows very closely that of the field stars. This
can be explained if these objects formed in a highly biased set of
gas-rich proto-galaxies at high-redshift \citep{kravtsov}.

The second interesting property is that the satellites are quite
anisotropically distributed \citep{hartwick,kroupa,metz07}. This can
be reconciled with cold dark matter models if these galaxies have
fallen in from a few directions, perhaps in groups \citep{li-helmi},
or along filaments \citep{libeskind,zentner05,kang}. It is unclear
whether this peculiarity will remain once the sky coverage is more
complete, and once a thorough survey of the faintest objects is
available. Furthermore, establishing the reality of the dynamical
families proposed by \cite{lb2}, is only possible with knowledge of
the satellite's orbits. For this, precise tangential velocities
(i.e. proper motions) are required. Unfortunately only a few satellite
galaxies have proper motion measurements, mostly obtained using the
Hubble Space Telescope \citep[see e.g.][and references
therein]{piatek} and these are still affected by large
uncertainties. However, in the coming decade, the situation is going
to improve significantly thanks to Gaia.

In contrast, the dynamics of the (inner) halo globular cluster
population is much better known. For example, \cite{dinescu99} has
compiled a sample with $\ge 40$ clusters with well-measured tangential
velocities. The measurements are relatively modern (i.e. made with
high-precision measuring machines, such as used in the Southern Proper
Motion survey), and are tied to an absolute reference frame using,
e.g.  galaxies and QSOs, or stars with known absolute proper motion
from, e.g.  the Lick Northern Proper Motion Program or Hipparcos
\citep[for more details, see e.g.][and references therein]{dinescu99}.

The reliable measurements of the dynamics of globular clusters in
combination with dedicated programs to characterize their properties,
such as chemical abundance patterns, ages, horizontal-branch
morphology, and structural parameters has prompted the search for
correlations in the hope to trace Galactic history \citep[e.g.][and
references therein]{ashman-zepf,mackey-gilmore}. This is much in the
same spirit of the original work of \cite{sz} and \cite{zinn93}.

A recent highlight is the discovery of multiple stellar populations in
the most massive globular clusters \citep{lee99}. One proposed
explanation \citep[but not the only one, c.f.][and references
therein]{parmentier} is that these might be remaining cores of
disrupted nucleated dwarf galaxies, as is the case for M54 which is
located in the Sagittarius dwarf itself \citep{layden-ata}. These
peculiar globular clusters also show an extended horizontal branch
(EHB), and the presence of double or broadened main sequences
\citep{piotto07}.  In fact, about 25\% of the globular clusters in our
Galaxy exhibit unusual EHBs, which has prompted the idea that they
might have a unique origin. Furthermore, these globular clusters are
on average more massive and share the random kinematics characteristic
of normal globular clusters in the outer halo \citep{lee-gc}.

These results, as well as the evidence of dynamical relation between
the Sagittarius dwarf and six Galactic globular clusters
\citep{Bellazini}, suggest an accretion origin for the outer halo
clusters. This is also supported by the work of \cite{mackey-gilmore},
who conclude that all young halo clusters (with $\feh < -0.8$ dex and
red HB morphology, more common in the outer halo) and $\sim 15\%$ of
the old halo clusters ($\feh < -0.8$ dex and blue HB morphology) are
of external origin.

\begin{figure}
\includegraphics[width=1.0\textwidth,viewport=1 1 560 350,clip]{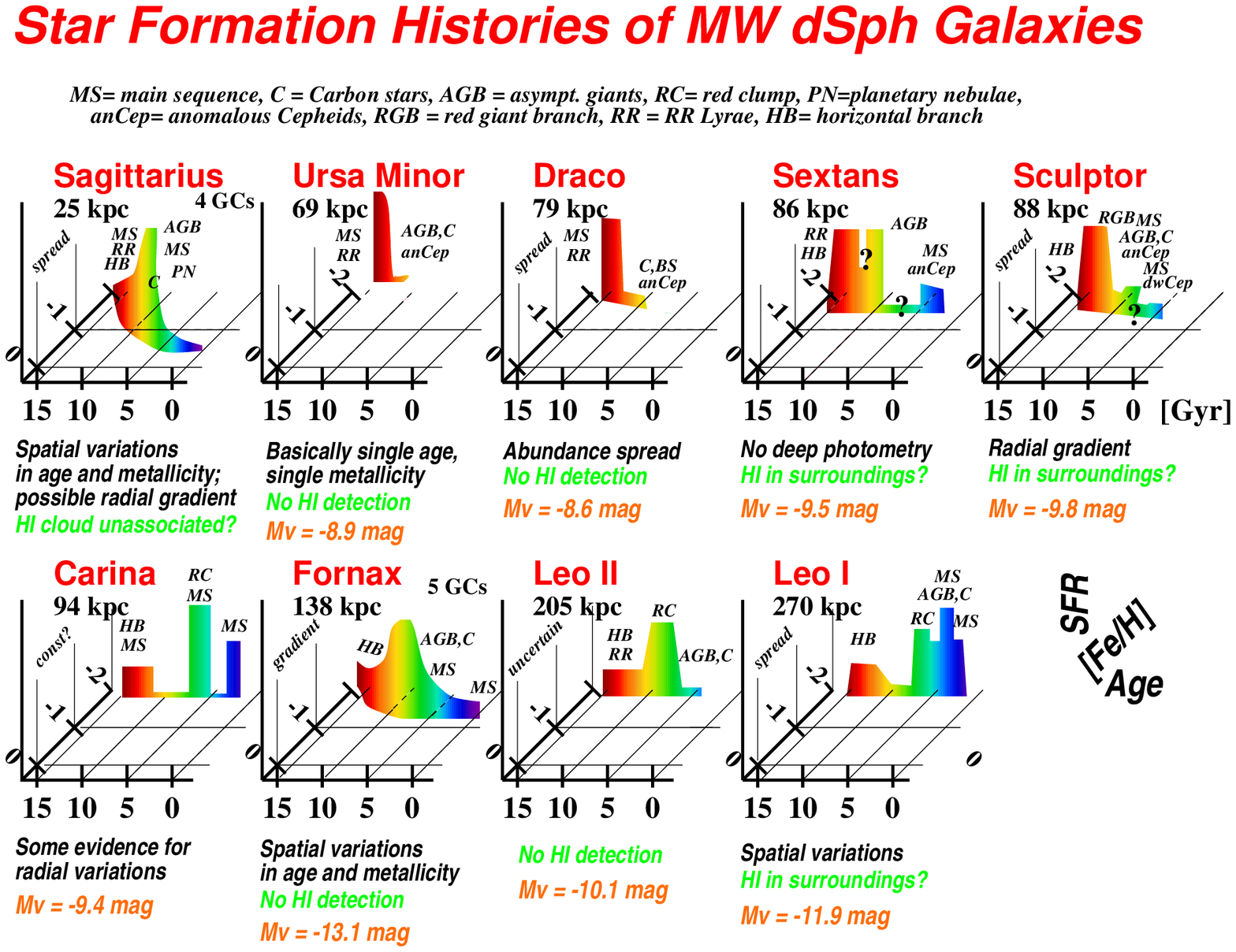}
\caption{Star formation histories of Milky Way dwarf spheroidal
satellites.  Each population box gives a schematic representation of
star formation rate (SFR) as a function of age and metallicity. [From
\cite{grebel}. Courtesy of Eva Grebel, and Fabio Favata (editor).] }
\label{fig:dSph-histories}
\end{figure}

Dwarf galaxies show, of course, much more complex stellar populations.
Furthermore, these galaxies depict a large variety in star formation
and chemical evolution histories \citep{grebel,dolphin}, as shown in
Figure \ref{fig:dSph-histories}. Some dwarf galaxies appear to have
stopped forming stars very early on, although it is clearly difficult
to disentangle whether this happened at $z\sim 10$, i.e. before
reionization, or at $z \sim 3 - 5$, i.e. after reionization
\citep{gnedin-n}. On the other hand, some dwarf galaxies have managed
to form stars until very recently, such as Fornax \citep{stetson} or
Sagittarius \citep{siegel07}.

There is, however, a remarkable uniformity in that all dwarf
spheroidals studied so far show an absence of stars with metallicities
below a value of $\feh \sim -3$~dex \citep{helmi06b}. This could
suggest a uniform pre-enrichment across the Local Group at high
redshift, before the first generations of stars in these objects
formed.

We are also interested in understanding the link between the field
stars in the halo and those still bound to the satellites of the Milky
Way \citep{sales}. For example, the contribution of present-day dwarf
galaxies to the halo near the Sun can be ruled out on the basis of the
colors and chemical properties of the stellar populations
\citep{unavane,venn04}.  These dwarfs have had a Hubble time to evolve
in relative isolation, and it is reasonable that their stellar
populations would not resemble those of objects disrupted a long time
ago. However, one may wonder what the relation is between the
progenitors of the dwarf galaxies and the halo building blocks.  Under
the assumption that the very metal poor stars represent the first
generations of stars formed, the apparent lack of such objects in the
dwarf spheroidals compared to the halo, implies that their progenitors
were fundamentally different. One possible scenario is that the halo
building blocks were associated to high-density peaks that collapsed
at higher redshifts, while the dwarf galaxies presumably descend from
more average density fluctuations in the early Universe
\citep{helmi06b,salvadori08}.

Nevertheless we know at least of one example of a significant
contributor to the outer halo: the Sagittarius dwarf galaxy. Although
it is currently unclear how massive this object was initially, current
models postulate an initial stellar mass of $\sim 2 - 5 \times 10^8
\sm$ \citep{helmi04a,law}, which is slightly smaller but comparable to
the total mass of the stellar halo obtained by extrapolating the local
density and assuming a power-law distribution as described in
Section~\ref{sec:struct}. Therefore, given that we are just beginning to
map the outer stellar halo in sufficient detail, it is very likely
that much of this region of the Galaxy is the result of the accretion
of dwarf galaxies like those we observe today.

\section{Formation scenarios}
\label{sec:formation}

There has been a tendency in the literature to classify formation
scenarios according to two traditional models introduced in the 1960s
and 1970s respectively, to describe the evolution of the Galaxy.

The first model was proposed by \cite{els}, where ``the oldest stars
were formed out of gas falling toward the galactic center in the
radial direction and collapsing from the halo onto the plane. The
collapse was very rapid and only a few times $10^8$ years were
required for the gas to attain circular orbits in equilibrium''. The
influence of this work has been phenomenal, not just on models of the
stellar halo, but also in the field of galaxy evolution in
general. The second model, which is often used as a counter scenario,
is that by \cite{sz}.  On the basis of the lack of an abundance
gradient for the outer halo globular clusters, these authors proposed
that ``these clusters formed in a number of small protogalaxies that
subsequently merged to form the present Galactic halo''. For many
authors, this latter model is more reminiscent of the currently
popular hierarchical paradigm.  However, as we shall see below,
ingredients of both models are likely to be present in the assembly of
a Galactic system in a hierarchical cosmology.

It is therefore perhaps more instructive to distinguish models
according to the degree of dissipation involved in the formation of a
given component. In this context, the \cite{els} model would fit in a
category where the large majority of the nearby halo stars were formed
during a dissipative process (namely the collapse of a gaseous
cloud). On the other hand, in the \cite{sz} model, the assembly of the
outer halo would have proceeded without any dissipation.

\subsection{Global properties}
\label{sec:formation-global}

Several authors have attempted to model the formation and
characteristics of the Galactic stellar halo. Different techniques
have been applied, such as numerical simulations with and without gas
physics and star formation, which are sometimes (fully) embedded in a
cosmological context. Other attempts include the use of
phenomenological descriptions of the evolution of baryons inside
halos, usually in combination with an N-body simulation that provides
the dynamical history of the system in the context of the hierarchical
paradigm.

The global properties of the Galactic stellar halo, such as its shape,
correlations with kinematics and metallicity distribution, have been
reproduced in many of these works. 

For example, \cite{samland} have modeled the formation of a disk
galaxy in a slowly growing dark matter halo. In essence, this model is
quite similar to the \cite{els} scenario, but takes into consideration
the fact that galaxies have grown in mass from the Big Bang until the
present time. They follow the cooling of gas, star formation and
feedback processes in great detail, and are able to obtain a realistic
disk galaxy at the present time. In this simulation, the stellar halo
defined by the stars with [Fe/H]$< -1.9$ dex is not rotating, while
the more metal-rich stars with $-1.9 < \feh <-0.85$ are rotating at
$\sim 70$ \kms. The first component is slightly flattened in the same
sense as the disk, while the second component is significantly flatter
and largely supported by rotation. In this sense, one may argue that
it resembles more closely a thick disk than the inner halo of the
Milky Way. 

More realistic attempts to simulate the formation of a disk galaxy
like the Milky Way in a cosmological framework include those of
\cite{steinmetz-m,bekki-c} and \cite{brook03}. These authors have
modeled the collapse of an isolated, rigidly rotating sphere of $\sim
10^{12} \sm$ onto which small-scale fluctuations according to a cold
dark matter spectrum have been imposed. In these works the evolution
of the dark matter and of the gas, star formation, etc.\ are followed
self-consistently.  In this sense they are, in principle, able to
determine the relative importance of dissipative processes versus
dissipationless mergers in the build-up of the halo. However, the
numerical resolution in these studies is low (the spatial resolution,
as measured by the softening, is typically $\sim 2.5$ kpc), and the
number of particles is small. For example, in \cite{brook03} the
dominant disk component and the spheroid at the final time are
represented by $\sim 4 \times 10^4$ star particles. This implies that
the resolution is not sufficient to determine the detailed properties
of the simulated galaxies robustly (e.g. small scale substructure such
as dwarf galaxies and streams are not well resolved). Nevertheless,
global (average) properties are likely to be reliable. For example,
\cite{steinmetz-m} find in their simulations that the stellar halo
(defined as the more metal-poor component) is triaxial in shape
($c/a=0.65$ and $b/a = 0.85$, with the minor axis roughly aligned with
the angular momentum of the disk), and that the density profile is
well-fit by a de Vaucouleurs law.  In the simulation of \cite{bekki-c}
the outer halo is completely built up through accretion, while the
inner halo results from merging between gas rich clumps. The inner
halo is also adiabatically contracted by the formation of the disk
\citep{binney-may,chiba-beers01}, which explains its more-flattened
shape. In agreement with \cite{steinmetz-m}, \cite{bekki-c} find that
the density profile varies from $r^{-3}$ at $r < 10$ kpc to $r^{-4}$
at larger distances.

This is also in agreement with the simulations of
\cite{abadi03a}. These authors have run fully cosmological
hydrodynamical simulations of the formation of a disk galaxy. Although
their final galaxies are of too early Hubble type (most of them have
bulges that are too dominant compared to the disk component), they
find some common (and hence presumably general) characteristics.  The
surface brightness profile of the outer halo components of these
galaxies ($r \ge 20$ kpc) can be fit with a Sersic law, with $\langle
n \rangle = 6.3$ and $\langle R_{\rm eff} \rangle = 7.7 $ kpc,
i.e. $R^{-2.3}$ at $\sim 20$ kpc and $R^{-3.5}$ around the virial
radius. The shape of the outer halos is very mildly triaxial $\langle
b/a \rangle = 0.91$ and $\langle c/a \rangle = 0.84$. These outer
regions ($r \ge 20$ kpc) contain mostly accreted stars (with fractions
greater than 90\%), which originate in the most massive mergers that
the central galaxy has experienced over its lifetime (the contribution
of small accretion events may be underestimated because of resolution,
but this is nevertheless unlikely to be the major channel by which
galaxies grow).  The inner halos are formed {\it in-situ} and are much
more flattened $c/a \sim 0.62$ \citep{abadi06}.

\cite{diemand} provide an alternative view from dark matter only
cosmological simulations. Following \cite{moore2001}, they identify
the most bound particles in halos that have collapsed at very high
redshift (i.e. 2.5$\sigma$ peaks at $z=12$) and find that these end up
in the inner regions of the most massive object at the present time.
These particles (which are meant to represent stars) are distributed
according to a very steep density profile $r^{n}$ with index $n \sim
-3$ in analogy to the stars in the stellar halo of our Galaxy.

A complementary approach has been developed by \cite{bj05} who follow
the formation of the stellar halo purely built via accretion of
satellites. The evolution of the baryonic components of the satellites
(fraction of cold and hot gas, star formation and chemical enrichment
histories) are followed through a phenomenological semi-analytic
approach \citep[as in][]{wf,kwg,somerville}. The satellites are placed on
orbits (consistent with those found in cosmological simulations,
e.g. \cite{benson05}), around a slowly growing gravitational potential
that models the Galaxy (resembling the approach of
\cite{samland}). In this sense, they are limited to times after which
the central potential has settled (i.e after major mergers), and so,
essentially they are well-placed to tackle the formation of the outer
halo, i.e. $r > 15-20$ kpc. Through this modeling they find that the
density profile of their stellar halos vary from $\rho \propto r^{-2}$
at 20 kpc to $r^{-4}$ beyond 50 kpc.

The higher resolution and more detailed modeling of \cite{bj05} also
allows them to make predictions about the distribution of chemical
abundances of the Milky Way outer halo stars. \cite{font06} find that
the mean metallicity of the halo depends on the accretion history, and
that gradients in metallicity of $\sim 0.5$ dex are not necessarily
uncommon in their accreted halos. They predict that the stars in the
outer halo should have lower [$\alpha$/Fe] than stars in the inner
Galaxy. In their models, the most metal-poor stars are located in the
inner 10 kpc, which at face value is opposite to the trend suggested
by \cite{carollo}.

\begin{figure}
\includegraphics[width=0.5\textwidth]{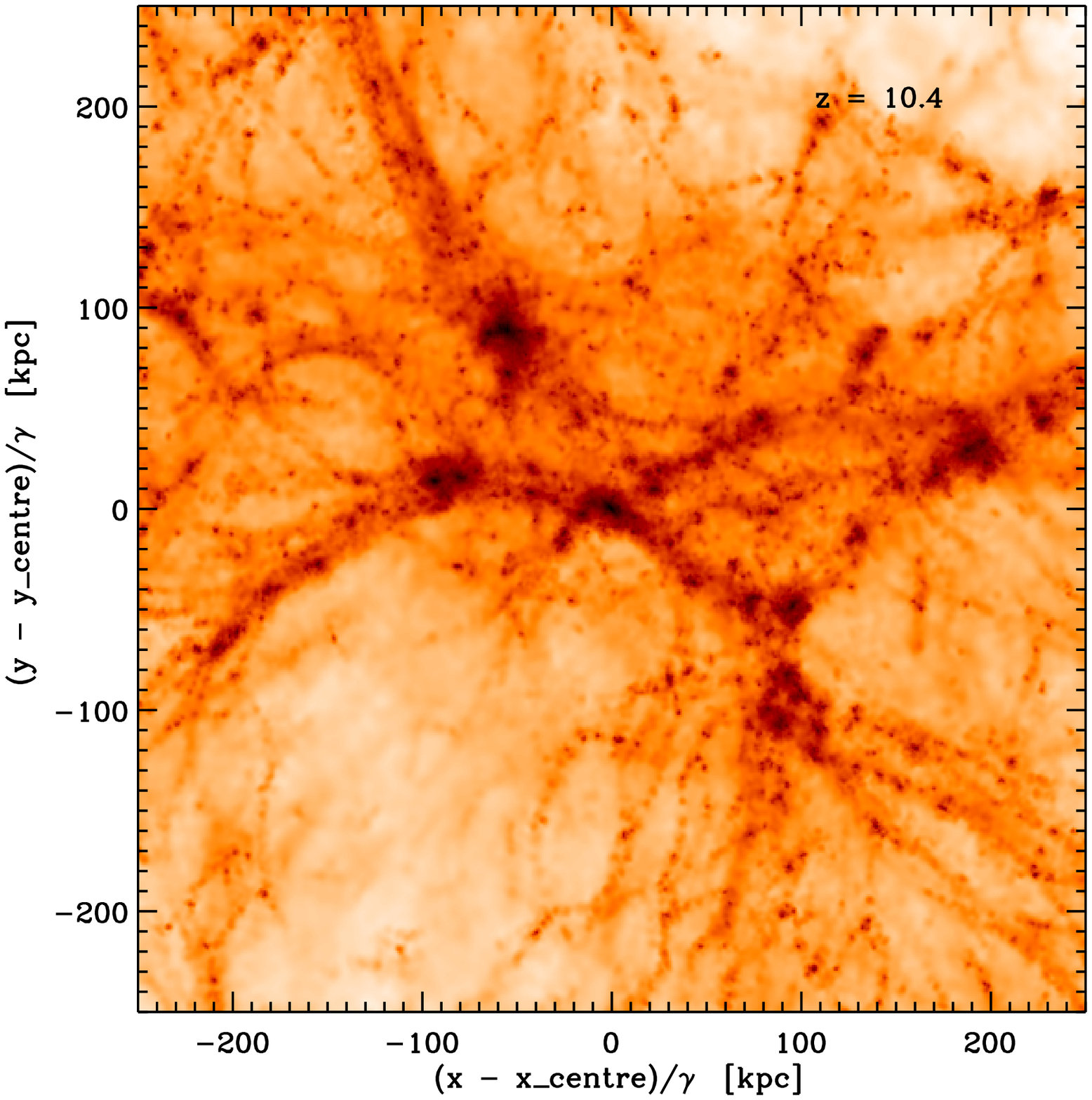}
\includegraphics[width=0.5\textwidth]{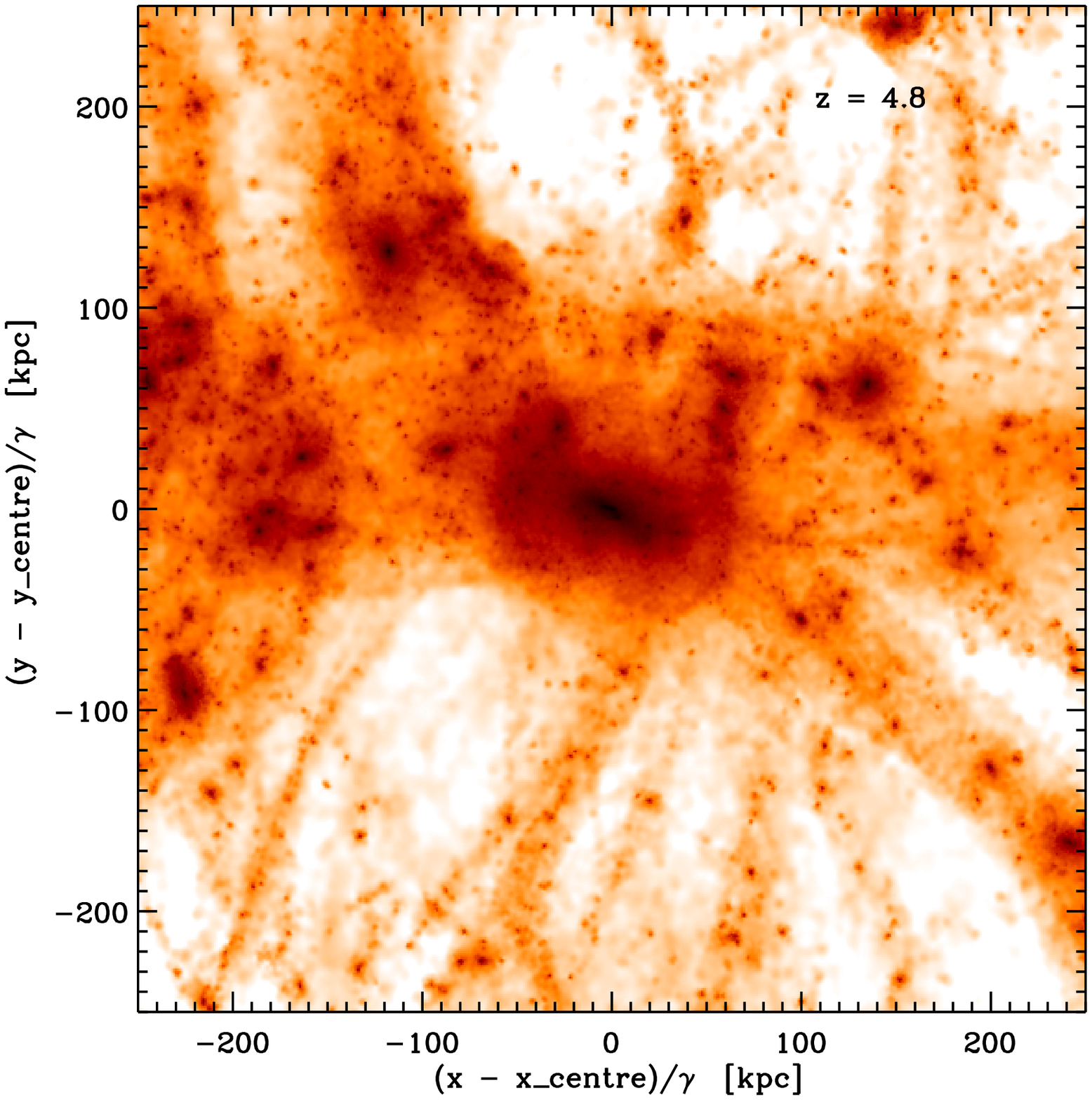}
\includegraphics[width=0.5\textwidth]{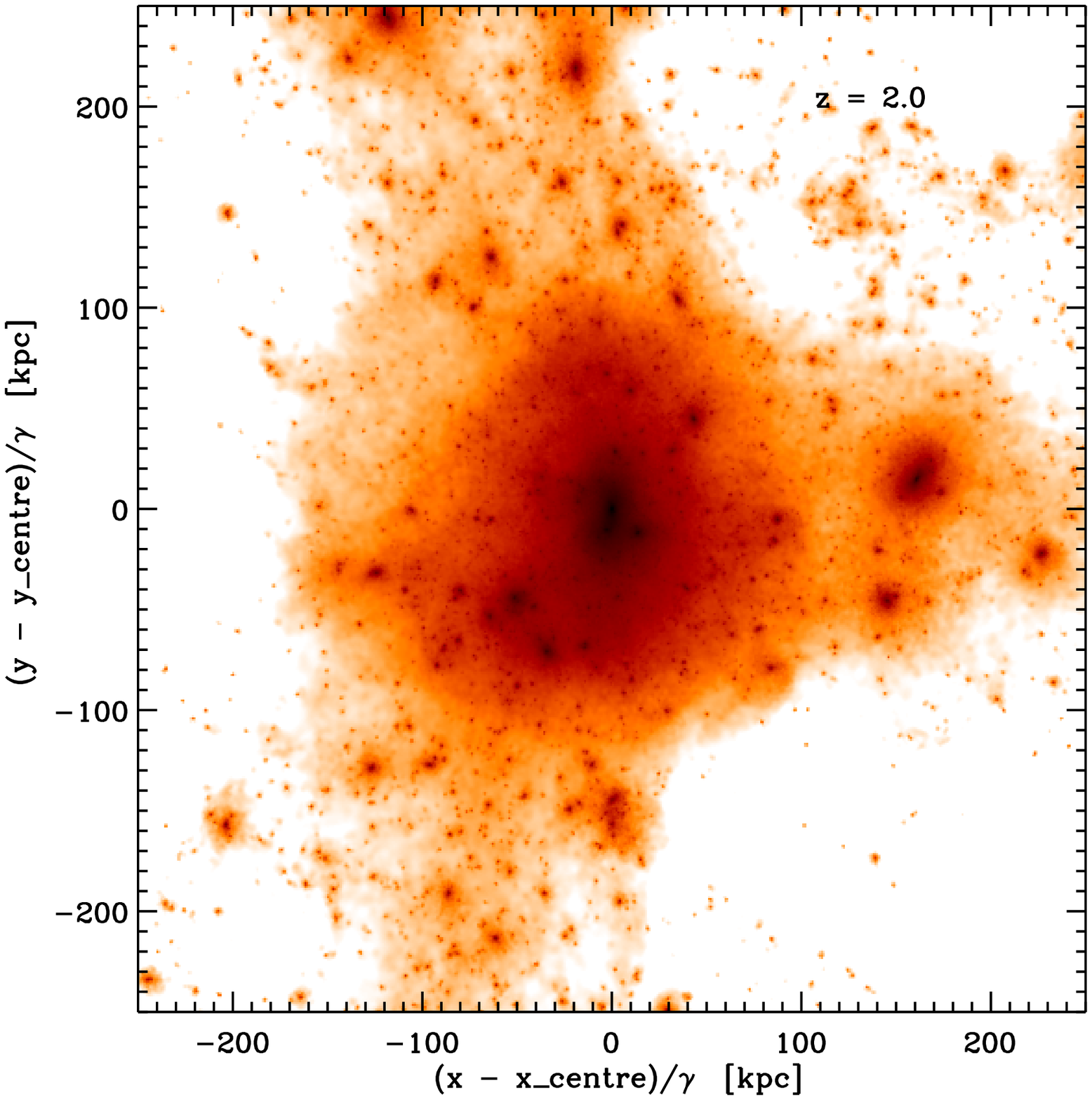}
\includegraphics[width=0.5\textwidth]{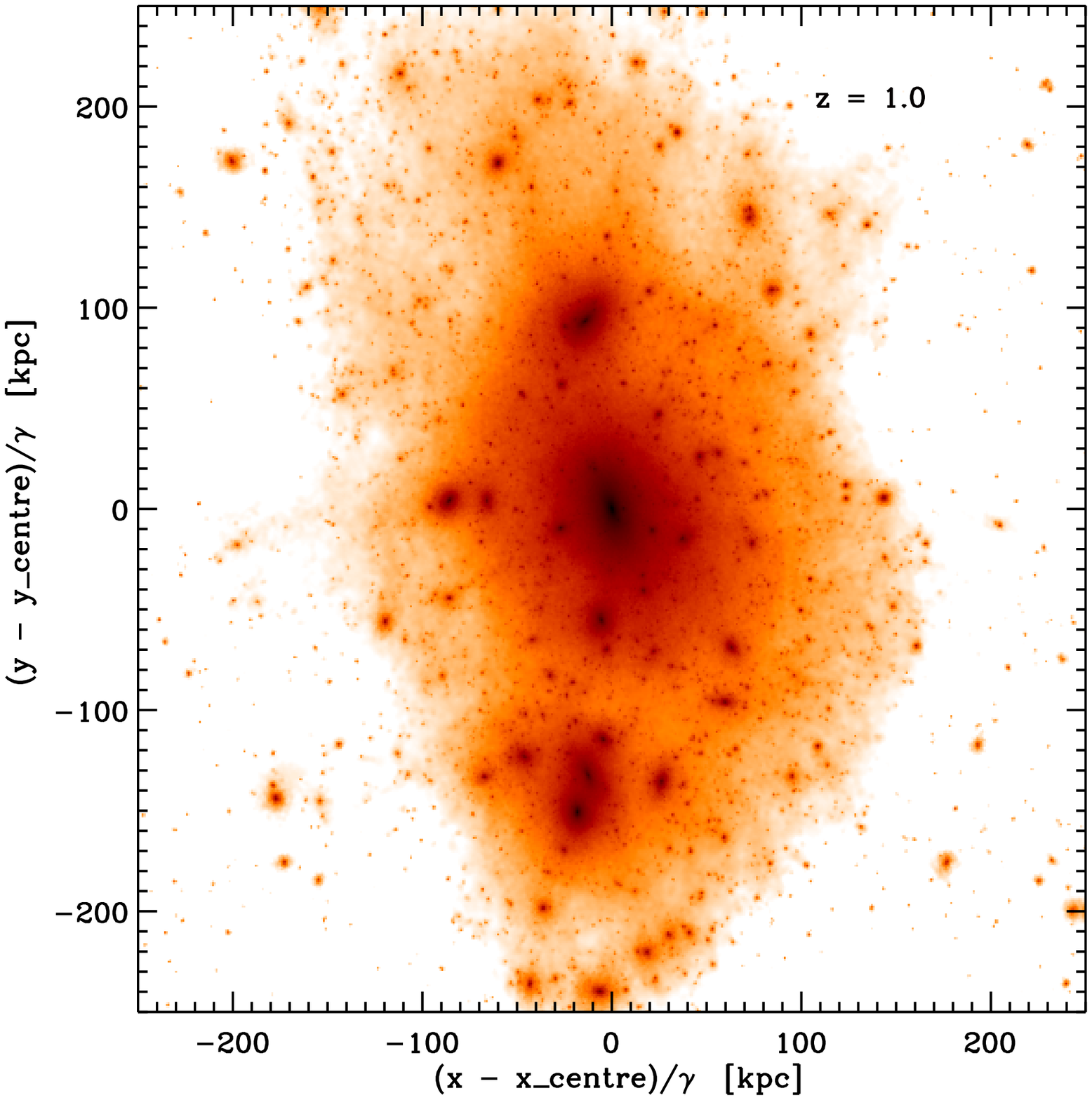}
\includegraphics[width=0.5\textwidth]{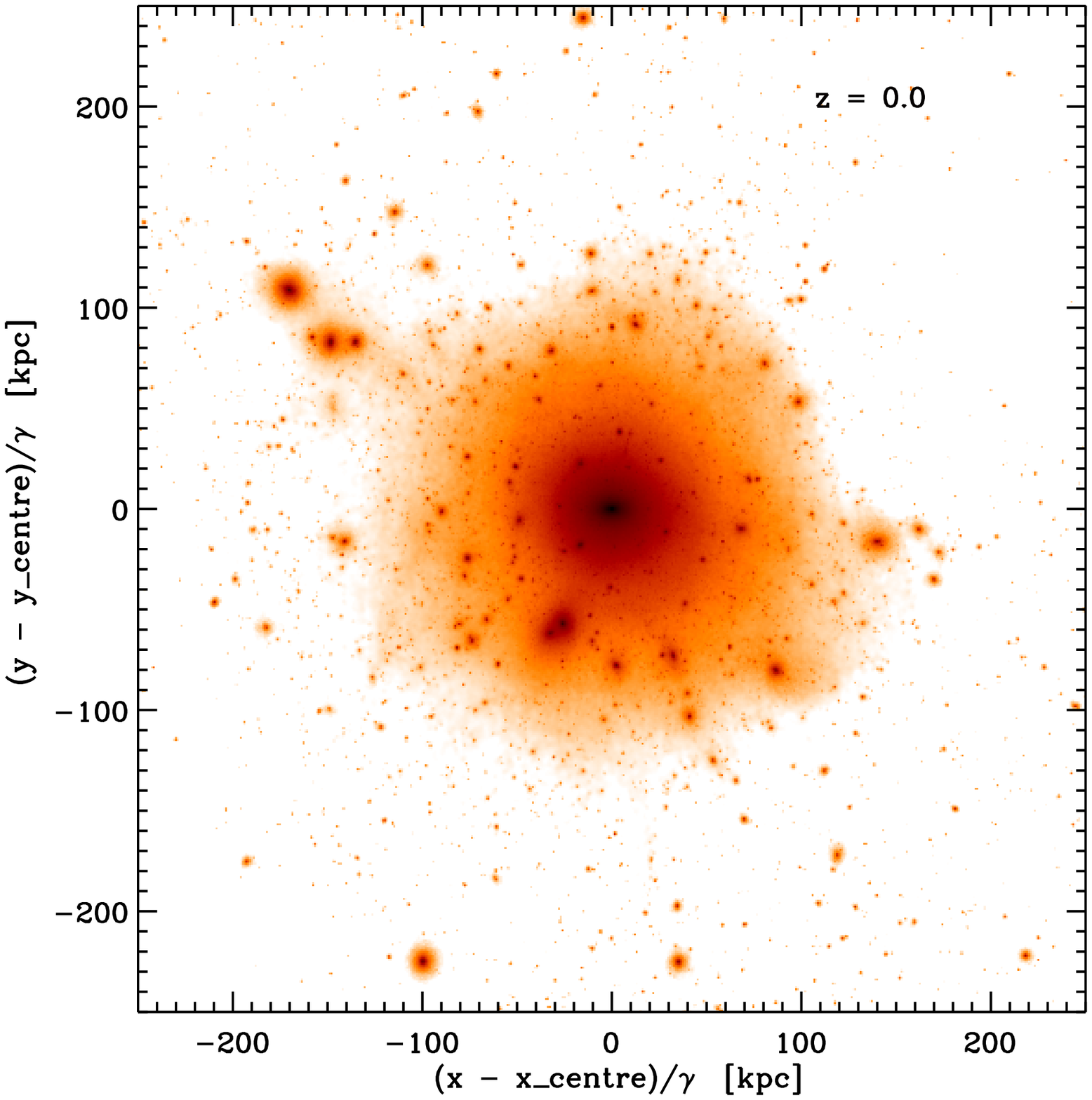}
\includegraphics[width=0.5\textwidth]{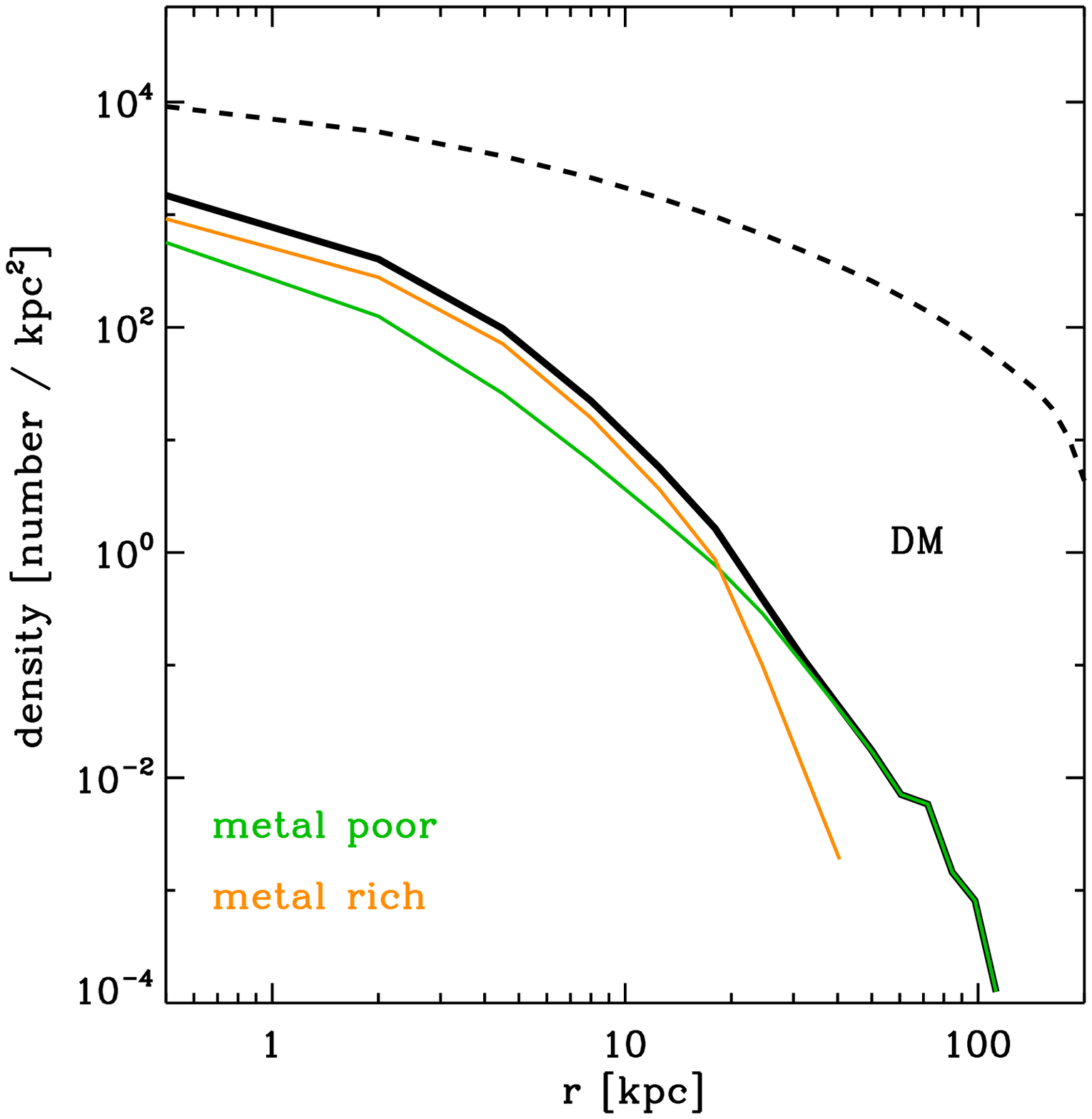}
\caption{Evolution of the dark matter distribution in a
high-resolution cosmological simulation of the formation of a Milky
Way-like halo by \cite{stoehr}. By identifying the cores of dark
matter substructures with luminous satellites, \cite{gdl} can follow
the formation and evolution of a spheroidal component, which would
correspond to the stellar halo. For example, by analyzing where these
objects deposit their debris, they can study the link between the
chemistry and the structure of the stellar halo. This is shown in the
bottom right panel, where the green and orange curves denote the
surface density profile of metal-poor and metal-rich ``stars'',
respectively. In comparison, the dashed black curve corresponds to the
dark matter halo, which shows a significantly more shallow and
extended distribution.[From \cite{gdl}. Courtesy of Gabriella de
Lucia.]}
\label{fig:gdl}
\end{figure}

The hybrid cosmological approach of \cite{gdl} combines elements of
the work of \cite{diemand} and \cite{bj05}. These authors use a
high-resolution cosmological simulation of the formation of a Milky
Way-sized dark matter halo coupled to a semi-analytic galaxy formation
code, to obtain insights into the global evolution of the Galaxy. By
identifying the most bound particles of substructures in the
simulation with the luminous cores of satellite galaxies, they are
able to follow the structure but also the metallicity and age
distribution of the ``simulated'' stellar halo. As shown in Figure
\ref{fig:gdl}, also \cite{gdl} find that the stellar halo is much more
centrally concentrated than the dark matter halo. Furthermore, their
inner halo is dominated by higher metallicity ``stars'', {\it
vis-a-vis} the outer halo, with the transition occurring at $\sim 10 -
15$ kpc from the center, reminiscent of the findings of
\cite{carollo}.

\subsection{Current status}

The discussion above shows that it is possible to reproduce the
density profile and shape of the inner stellar halo in a model in
which all stars were formed {\it in situ} as in \cite{samland}, as
well as in a model in which all stars were accreted as in
\cite{diemand}, \cite{bj05} and \cite{gdl}. Furthermore, in all the
models discussed here the oldest stars are located in the spheroid
(halo+bulge), which implies that their age distribution and the broad
chemical properties (low mean metallicity and $\alpha$-enhancement)
are also likely to be reproduced. Nevertheless, much more work is
required to understand the expected trends as well as predictions for
the detailed chemical abundance patterns that have now been observed
in such exquisite detail for nearby halo stars. The work of
\cite{chiappini} is an excellent example of how this can be done, but
it needs to be embedded in a more cosmological framework.

The discussion in the previous paragraphs also shows that we do not
really possess a clear view of what are the exact predictions of the
hierarchical paradigm on the scales of the Milky Way and its
satellites. The relatively simplistic and coarse fully cosmological
simulations are limited in their predictive power at the level of
detail that is required for the Milky Way galaxy.  The semi-analytic
approach of combining high-resolution dark matter simulations with
phenomenological descriptions of the evolution of the baryonic
components, is a powerful and promising technique, that will be
developed much further in the near future \citep[see e.g.][]{gdl}. The
first attempts by \cite{bj05} at modeling the outer halo have yielded
extremely interesting results, despite the fact that the growth of the
Galaxy is followed under rather ideal conditions.

From the observational point of view, there seems to be clear
consensus that the outer halo is very lumpy. This would imply that
there is little room for a component predominantly formed via a
dissipative collapse.

\begin{figure}
\includegraphics[width=0.7\textwidth,angle=270]{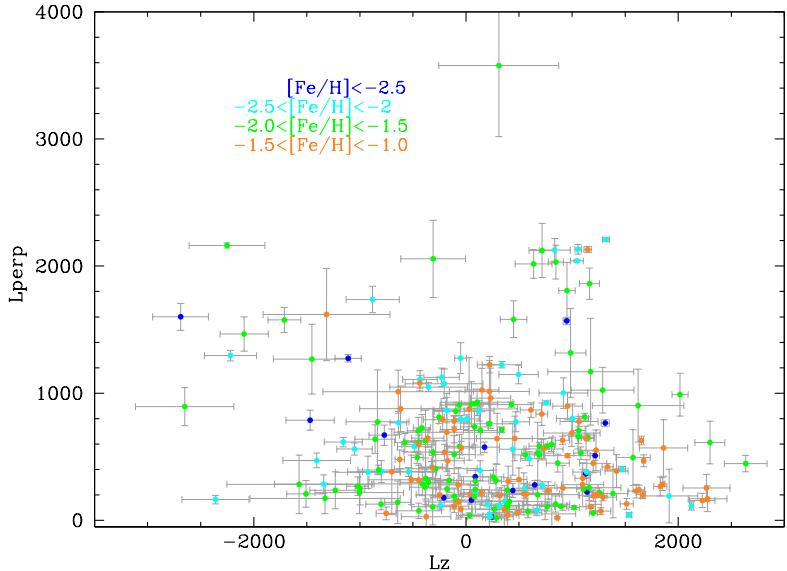}
\caption{Distribution of angular momentum for a sample of nearby halo
stars. The color-coding indicates different metallicity ranges. Stars
with orbits confined to the plane (${\rm Lperp} = \sqrt{L_x^2 + L_y^2}
< 350$ kpc\kms) tend to have a net prograde motion (corresponding to
$V_\phi \sim 50$\kms), and their distribution is very smooth in this
space. Stars with highly-inclined orbits (${\rm Lperp} > 350$ kpc\kms)
have no net rotation and their distribution is quite lumpy, indicative
of the presence of merger debris. It is also interesting to note that
this ``higher halo'' is more metal-poor on average. [From
\cite{morrison07}. Courtesy of Heather Morrison.]}
\label{fig:hlm}
\end{figure}

In the case of the inner stellar halo, the situation is much less
clear.  Recently, \cite{morrison07} analyzed the distribution of
angular momentum for a sample of nearby metal-poor halo stars with
accurate velocities. These authors found a break in the properties of
these stars depending on their angular momentum as shown in Figure
\ref{fig:hlm}. The distribution of angular momentum is clearly not
smooth, in particular for the higher halo (stars on highly inclined
orbits), which suggests some amount of accretion in the build-up of
this component. 

Perhaps the most direct way of testing the amount of ``dry''
(i.e. dissipationless) accretion or mergers that have contributed to
the stellar halo near the Sun, is to find the debris streams that are
expected to be present. The granularity in the velocity distribution
of stars should be measurable, if present, in large catalogs of halo
stars ($ > 5000$) with very accurate kinematics (errors $\le
10$\kms). The only processes by which these streams are expected to
diffuse away are through collisions, or if the gravitational potential
is (or has been) extremely chaotic. Galaxies are essentially
collisionless systems \citep{bt}, and the effect of dark matter
substructure on the coherence of tidal streams has been shown to be
very small \citep{kvj02}. Therefore one may only need to worry about
the chaotic nature of the potential. First studies performed on fully
cosmological dark matter simulations show evidence that the number of
stellar streams expected near the Sun is the same as in steady-state
models of the accretion of satellite galaxies onto a fixed
gravitational potential \citep{hws03}. The Galactic bar may also be a
source of orbital diffusion. This triaxial potential is known to
induce resonances amongst stars on low inclination orbits
\citep{dehnen,famaey}, and may well cause streams to become more
diffuse in time, and hence harder to recover.

Another promising approach is that of chemical tagging, i.e. stars
born within the same molecular cloud should have very similar chemical
abundances as shown by \cite{desilva}. We have yet to understand how
much scatter there can be in a galaxy that was accreted very early on,
and in particular, if similar degrees of coherence in the abundances
are expected for such (much larger) systems. For example, dwarf
galaxies today show metallicity spreads of the order of 0.5 dex
\citep{mateo-rv}. Although one expects stars in a stream to come from
a small region of phase-space, it is unclear how different their
chemical abundances may be. For example, a satellite that has been
disrupted only relatively recently will give rise to streams that are
less chemically homogeneous than one whose streams are dynamically
older. In the first case, the metallicity spread will likely reflect
that of the whole galaxy, while in the second case it may reflect only
that of a small portion of the system. In the limit of an infinite
amount of time, the stars in a given location on a stream should have
all been born within the same molecular cloud, and hence have the same
exact abundances, provided diffusion processes are unimportant.

An accurate measurement of the ages of stars in the halo would yield
unique insights. This would give a very robust test of the degree of
dissipation involved in the formation of the stellar halo. For
example, if the stars were formed predominantly during gas-rich
mergers, the star formation history of the halo would have been
largely bursty. In this case, each burst would be associated to each
one of those gas-rich merger events. If on the other hand, the halo
formed only via the accretion of a few objects containing some stars,
each of these objects will have had a different star formation
history, and hence a larger spread in ages might be expected. Note
that in the limit in which gas-rich mergers have taken place
successively on a very short timescale, such a model will be
essentially indistinguishable from the original \cite{els} proposal
for the formation of the inner halo, even if the $\Lambda$CDM model is
correct.

\section{Halos in other galaxies}
\label{sec:ext-halos}

Stellar halos are generally such low surface brightness components of
galaxies that detecting these beyond the Milky Way is a major
challenge. For example, \cite{morrison93} estimate the surface
brightness of the Galactic halo at the solar radius to be $\Sigma_V
\sim 27.7$ mag/arcsec$^2$.

\begin{figure}
\includegraphics[width=1.0\textwidth]{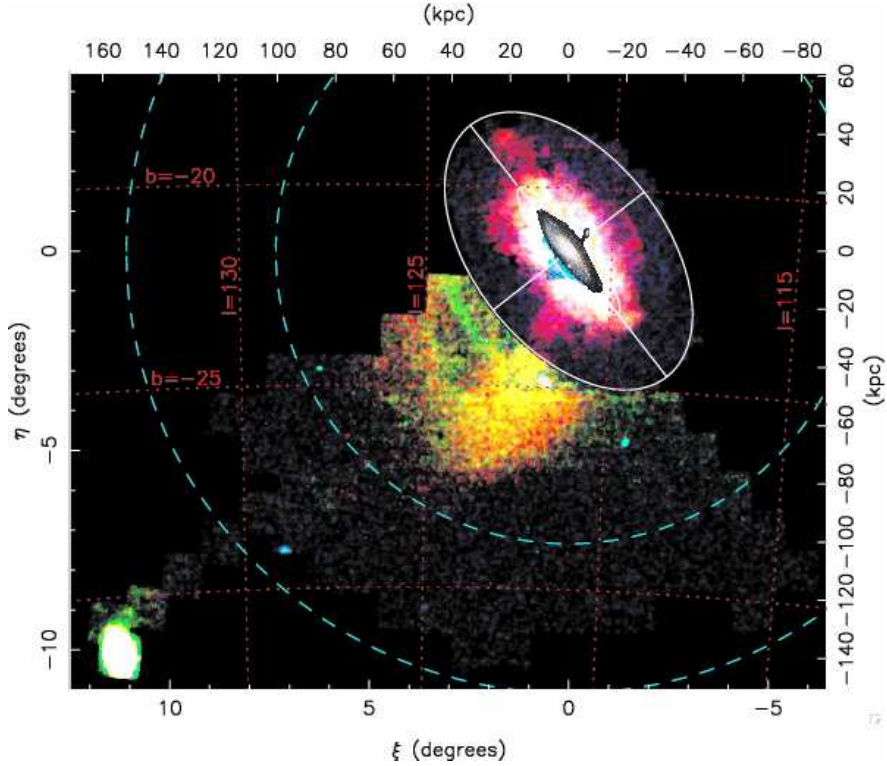}
\caption{View of the halo of M31. Red, green and blue show,
respectively, stars with $-0.7 < \feh < 0$, $-1.7 < \feh < -0.7$ and
$-3.0 < \feh < -1.7$. The differences in stellar populations between
the Giant Stream and the several minor axis streams can be seen as
striking differences in color.  At the center there is a scaled image
of the central of M31.  [From \cite{ibata07}. Courtesy of Rodrigo
Ibata. Reproduced with permission of the AAS.]}
\label{fig:m31}
\end{figure}

Direct testimony of the difficulty of such an enterprise is given by
M31.  Its spheroid was long believed to be a simple extension of its
bulge, since ``metal-rich'' stars were found even out to a radius of
30 kpc following a de Vaucouleurs profile
\citep[e.g.][]{mould,brown}. However, \cite{kalirai} and
\cite{chapman} recently discovered a metal-poor halo out to 160 kpc
that is similar to that of the Galaxy, both in terms of its
metallicity, radial profile, as well as the amount of
substructure. Our external point of view has allowed the study of this
halo in a much more global fashion \citep{ibata07}, and has provided
spectacular images as shown in Figure~\ref{fig:m31}. Just as for the
Milky Way, in the halo of M31 we find a giant stream as well as a
wealth of fainter substructures at surface brightness levels of $\sim
30$~mag/arcsec$^2$.

M33 seems to have a halo much like that of the Milky Way and M31, with
a similar metallicity $\feh \sim -1.6$ dex, despite its smaller total
mass \citep{mcc}. There is also some evidence that the Large
Magellanic Cloud may have a stellar halo \citep{minniti}. However, the
origin of this halo is less clear because the LMC appears to have been
tidally perturbed by the Milky Way \citep[its disk is asymmetric and
fairly thick as shown by][]{vdmarel01,vdmarel}. Such strong tides
could give rise to a halo-like component \citep{mayer}.

Beyond the Local Group, there have been a few studies of halos using
surface brightness photometry. This work is very challenging because
it requires characterizing a component several magnitudes below the
sky level. A few halos have been detected in this way (e.g. NGC5907,
NGC3115, see \cite{morrison-sackett,sackett}) as well as various tidal
features \citep{zheng,pohlen}. An interesting detection was obtained
by \cite{zibetti} who, stacking images of 1047 edge-on disk galaxies,
were able to measure a diffuse moderately flattened stellar halo, well
described by a power law $r^{-3}$ density profile.

An alternative, possibly more powerful approach is to study the
resolved stellar populations in the outskirts of (disk) galaxies, as
frequently done in the Local Group. Using the Advanced Camera for
Surveys (ACS) on HST \cite{mouhcine} and \cite{de-jong} have been able
to resolve stars a few magnitudes below the tip of the red giant
branch for a sample of nearby disk galaxies (located at distances
between 4 and 14 Mpc). For example, \cite{mouhcine} find that the
stellar halos are generally old, and typically more metal-poor than
the main galaxy. These authors also find large variety in their
properties. An example is NGC891, often referred to as the twin of the
Milky Way, whose halo contains a much larger spread in stellar
populations than is apparent for the Milky Way at a similar distance
\citep{mouhcine-891}. On the other hand, \cite{mouhcine05b} have
suggested that there is a correlation between the stellar halo
metallicity and the parent galaxy luminosity. Our Galaxy falls off
this metallicity-luminosity relation by more than 1 dex, i.e. the
Galactic stellar halo would be unusual, or too metal-poor \citep[see
also][for a similar statement]{hammer}.

Some caution is necessary when interpreting these observations,
especially for the following two reasons. First, in the absence of
spectroscopic measurements, the chemical properties of the stars
cannot be determined reliably because of the well-known
age-metallicity degeneracy (a clear example of which is the dwarf
galaxy Carina, \cite{koch}). Secondly, stellar halos can be very
lumpy, particularly at very low surface brightness levels, as shown in
Figures~\ref{fig:stellar-halo-acc} and \ref{fig:m31}. ACS/HST has a
small field of view, implying that the results are quite sensitive to
small-scale substructure. It is not unlikely, given what we know about
the halos of the Milky Way and M31, that such a field would be
dominated by just one stream such as for example the Sagittarius or
the Giant stream around Andromeda. It may be misleading to correlate
the properties of a dynamically young component such as the (outer)
stellar halo, with the global properties of the parent galaxy, until a
more global view of these halos is available \citep{ferguson}.

The avenue of mapping and resolving the stellar halos of nearby disk
galaxies is clearly a very worthwhile goal, and one that should be
pursued in the near future. The advent of Extremely Large Telescopes
(ELTs) should allow wide-field surveys of the halos of nearby
galaxies (also of earlier type), perhaps out to the Virgo cluster. A
significant addendum would be spectroscopy, both to determine
metallicities as well as the dynamics in the outskirts of these
objects.  In the hierarchical paradigm all galaxies should have
experienced mergers, and most frequently minor mergers because most of
the galaxies in the Universe are dwarfs. This implies that their
debris should end up in a stellar halo, and so this component is
expected to be ubiquitous. Therefore, unique insights into the
assembly histories of galaxies as well as on their dynamics can be
retrieved by studying their stellar halos.

\section{Epilogue}
\label{sec:concl}

The assembly process of a galaxy leaves imprints in the spatial,
kinematical, age and chemical distribution of its stars. We have
argued here that the Galactic halo should hold some of the best
preserved fossils of the formation history of our Galaxy.

A stellar halo may well be the most common galaxy component and a
natural outcome of the hierarchical build-up of galactic systems. For
example, mergers are expected to deposit debris in a spheroidal like
configuration. In the case of our Galaxy, both the halo field
population, as well as the globular clusters \citep{Cote} may have
such an origin, as evidenced by e.g. the Sgr dwarf \citep{Bellazini,
Dinescu}. Just like for stars, it is also possible that some fraction
of the globular clusters formed {\it in situ}, perhaps during early
gas-rich mergers that were presumably extremely common in the past
\citep{kravtsov}. By studying the dynamical properties of the globular
cluster population, and linking these to those of the field stars,
such formation scenarios can be tested
\citep[e.g.][]{casetti-dinescu}.

At present it is unclear what fraction of the Galaxy's stellar halo
has been built via dissipationless merging, and which by the
gravitational collapse of one or more gas clouds (i.e. via a
monolithic-like collapse or during gas rich mergers). There is
significant evidence of substructure in the form of streams in the
outer halo, but this is still at a rather qualitative level
\citep[although see][]{Bell}. It is also very likely that there is
currently a bias towards detecting the highest surface brightness
features, which presumably correspond to the most recent accretion
events of relatively massive satellites. Deep wide-field photometric
surveys should allow us to quantify the amount of substructure in the
outer halo in an unbiased manner, especially if accompanied by
spectroscopic surveys. Accurate photometry to enable quick and easy
identification of suitable halo tracers are a must. Planned surveys
such as VISTA\footnote{\sf http://www.vista.ac.uk}, the OmegaCam on
the VLT Survey Telescope, SkyMapper, PanStarrs\footnote{\sf
http://pan-starrs.ifa.hawaii.edu/public/} and the Large Synoptic
Survey Telescope (LSST\footnote{\sf http://www.lsst.org/}) will
significantly contribute to this enterprise. Wide-field imaging of
nearby disk galaxies as ongoing with HST \citep{de-jong}, and as
discussed for the ELTs, will allow us to establish the commonality of
such merger events.

The dynamical timescales are much shorter in the inner regions of
galaxies, which implies that halo substructures are no longer apparent
in physical space, but become very prominent in velocity space. For
example, a galaxy of the size of the Small Magellanic Cloud should
have contributed $\sim 50 - 100$ streams/cold moving groups near the
Sun if accreted roughly 10 Gyr ago \citep{hw99}. This implies that
full phase-space information is absolutely vital to recover the $\sim
500$ streams predicted to be present in the inner stellar halo if this
had been built up only via ``dry'' mergers. Current datasets are too
small, or only have a limited amount of sufficiently accurate
phase-space information to allow testing such a scenario. Of course
this implies that it is still possible that the halo formed largely
during a gas rich (merger) phase.

The definitive understanding of the structure and history of the
Galactic stellar halo is likely to be established only via a
``multi-dimensional'' approach. Knowledge of the stars kinematics
complemented with their chemical abundances and ages should allow us
to address questions such as:
\begin{enumerate}
\item What fraction of the halo has been built via accretion? How does
the importance of this process change with distance from the Galactic
center? 

\item When was the halo in place? 

\item How many accreted objects have contributed to build up our
Galaxy? Were these objects major contributors of cold gas, that later
produced the disk? Is there an evolutionary link between the disk and
halo?

\item What is the relation between the bulge and the stellar halo? The
bulge is as old as the halo but more metal-rich, and pressumably built
partly by a dynamical instability which led to the bar. Yet, a
fraction of its stars may have the same origin as some halo stars.

\item What were the properties of the accreted objects? What were
their masses, their star formation and chemical histories? How do they
compare to the small galaxies we see around us today?

\item What is the dynamical history of our Galaxy? How did its mass
and gravitational potential evolve in time?
\end{enumerate}

The answers to these and many other questions will all be within reach
in the coming decade, when large datasets with full phase-space and
astrophysical information become available, i.e.  in the era of Gaia.

\begin{acknowledgements}
This review is dedicated to Manuel and Mariano. Without their support
and their patience it would have never come to an end! I am grateful
to Michael Perryman for suggesting this review, and for giving me a
deadline (to try) to meet. NOVA, and NWO through the VIDI program are
acknowledged for financial support. Many friends and colleagues have
contributed to this review with thoughts, inspiration, figures and
feedback. All the figures are examples of this. I would like to thank
them and especially mention here Giuseppina Battaglia, Paul Harding,
Vanessa Hill, Mariano M\'endez, Heather Morrison, Michael Perryman,
Annie Robin, Laura Sales, Else Starkenburg and Eline Tolstoy. Dimitry
Rotstein and Norbert Christlieb are gratefully acknowledged for
bringing up some inaccuracies in the printed version of this review.
\end{acknowledgements}

% BibTeX users please use
%\bibliographystyle{spbasic}
%\bibliography{general}   % name your BibTeX data base

% Non-BibTeX users please use

\end{document}